\newif\iffinaldraft

\documentclass[a4paper,fleqn]{cas-sc}
\usepackage[authoryear]{natbib}

\usepackage{amsmath, amssymb}
\usepackage{bm}

\usepackage{graphicx}
\usepackage{tikz}

\usetikzlibrary{positioning, calc, arrows.meta}

\usepackage{hyperref}
\usepackage{url}
\usepackage{footmisc}

\usepackage{booktabs}
\usepackage{multirow}
\usepackage{makecell}
\usepackage{threeparttable}
\usepackage{adjustbox}

\usepackage[justification=justified,singlelinecheck=false]{caption}
\usepackage{subcaption}
\usepackage{pdflscape}
\usepackage{rotating}
\usepackage{float}
\usepackage[section]{placeins}
\makeatletter
\AtBeginDocument{%
 \expandafter\renewcommand\expandafter\subsection\expandafter{%
 \expandafter\@fb@secFB\subsection}%
}
\makeatother

\usepackage[dvipsnames]{xcolor}
\usepackage{soul}
\usepackage{xspace}
\usepackage{enumitem}
\usepackage{microtype}

\begin{document}

\let\WriteBookmarks\relax
\def\floatpagepagefraction{1}
\def\textpagefraction{.001}
 
\shorttitle{Behavioral Responses to E-Scooter Speed Governance}
 
\shortauthors{S. Choi, S. Yoo and S. Lee}

\title[mode=title]{Do E-Scooter Speed Governance Policies Reduce Harsh Acceleration and Deceleration? Evidence from 19.5 Million Trips Around a Regulatory Ban}

\author[1]{Seongjin Choi}[orcid=0000-0001-7140-537X]
\ead{chois@umn.edu}
\credit{Conceptualization of this study, Methodology, Software, Validation, Formal analysis, Writing - original draft, Writing - review and editing}

\author[2]{Sunbin Yoo}
\cormark[1]
\ead{sunbinyoo@skku.edu}
\credit{Conceptualization of this study, Methodology, Formal analysis, Writing - original draft}

\author[3]{Sugie Lee}
\credit{Conceptualization of this study, Methodology, Formal analysis, Writing - original draft}

\address[1]{Department of Civil, Environmental, and Geo- Engineering, University of Minnesota, Twin Cities, 500 Pillsbury Drive S.E., Minneapolis, MN 55455-0116, United States of America}
\address[2]{Department of Energy, Sungkyunkwan University, Suwon, Republic of Korea}
\address[3]{Department of Urban Engineering, Hanyang University, Seoul, Republic of Korea}

\cortext[cor1]{Corresponding author}

\begin{abstract}
Do e-scooter speed governance policies yield behavioral safety gains beyond the mechanical cap they impose? A firmware ceiling mechanically prevents speeding, but whether the same riders also generate fewer harsh accelerations and harsh decelerations when the ungoverned mode is withdrawn remains open. We analyze 19.5 million GPS-instrumented trips from 52 South Korean cities (February to November 2023). Our two-stage \textit{predict-then-validate} design targets two trip-level binary outcomes, any harsh-acceleration event and any harsh-deceleration event. In Phase~I, we predict each outcome's within-user reduction under an ungoverned-to-governed substitution, using a rider-heterogeneous random-parameters binary logit on the pre-ban period. In Phase~II, we validate these predictions using a difference-in-differences specification that exploits the operator's system-wide December~2023 removal of the ungoverned mode. The causal estimates confirm the Phase~I predictions in sign and order of magnitude on both outcomes, are Bonferroni-significant, and satisfy a 3-month pre-ban parallel-trends test. A within-user composition check finds no behavioral offsetting, indicating that firmware removal of an ungoverned mode lowers both harsh-event margins through a purely mechanical channel. These results imply that speed governance policies can deliver measurable safety gains on unconstrained behavioral margins.
\end{abstract}

\begin{keywords}
E-scooter safety\sep
Speed governance\sep
Harsh acceleration and deceleration\sep
Random-parameters logit\sep
Difference-in-differences\sep
Micromobility
\end{keywords}

\maketitle

\section{Introduction}
\label{sec:introduction}

E-scooter injuries have risen substantially alongside the expansion of dockless sharing services \citep{Aizpuru_2019,Hoveidaei_2025}. Emergency department studies report a consistent pattern, with head trauma, fractures, and soft-tissue wounds dominating \citep{Bloom_2021,Nellamattathil_2020}. Injury severity is closely tied to speed and trip characteristics \citep{Cicchino_2021,Yang_2020}. The power model of the speed--crash relationship predicts that a 10\% increase in mean speed raises injury crash frequency by approximately 20\%, with an especially steep gradient for vulnerable road users \citep{Elvik_2013,Elvik_2019}. Jurisdictions have responded by mandating speed caps (25~km/h in South Korea and France, 20~km/h in Germany, 15--20~mph across U.S.\ cities), but unlike motor-vehicle speed limits, these caps are enforced not through signage, radar, or penalties but through \textit{firmware speed governors} that electronically constrain motor output, making exceedance physically impossible. Early evaluations report mixed effects of such restrictions on injury incidence, with some instruments producing substantial reductions and others finding no measurable change \citep{Pakarinen_2023,Liukkonen_2023}.

However, whether speed governors produce safety gains \textit{beyond} the mechanical speed reduction itself remains an open question. The governor's effect on maximum speed is mechanical, since a firmware ceiling prevents exceedance by construction. What is less clear is whether the governed fleet also generates fewer harsh-kinematic events (harsh accelerations and harsh decelerations) and, symmetrically, whether riders compensate by riding more aggressively on the surviving governed modes. This concern is rooted in the risk-compensation literature \citep{Peltzman_1975,Wilde_1982}, though subsequent work suggests that compensation is unlikely when the intervention is involuntary and non-salient \citep{Hedlund_2000,Cohen_Einav_2003}. The existing micromobility literature does not resolve this question. Studies examine one outcome at a time, such as spot speeds, injury counts, or isolated kinematic traces \citep{Ma_2021,White_2023}, and policy evaluations observe only aggregate outcomes such as injury incidence \citep{Liukkonen_2023,Pakarinen_2023,Asensio_2022,Caggiani_2023}. No prior study has tested behavioral compensation across multiple unconstrained margins for any micromobility intervention \citep{Cho_2023,Tian_2024}.

To address these gaps, we exploit a natural experiment from Swing, a major South Korean shared e-scooter operator that offered governed and ungoverned modes on identical vehicle hardware across 52 cities. In December 2023, the operator removed the ungoverned mode system-wide, providing an exogenous shock. Our dataset comprises 19.5 million GPS-instrumented trips with per-point speed profiles from over one million riders (Section~\ref{sec:data}).

Our design is a two-stage predict-then-validate framework with four components.
\begin{itemize}[leftmargin=*, nosep]
 \item \textbf{Outcomes.} For each trip we record whether it contained any harsh acceleration event and whether it contained any harsh deceleration event (both binary $\{0,1\}$ indicators at the $\pm 0.5$~m/s$^2$ threshold). These two outcomes are what Phase~I predicts and Phase~II tests. Speeding ($\mathbf{1}[\max\text{speed}>25\text{ km/h}]$) is reported separately as a firmware-validation check, since the governor makes it mechanically impossible, not as a behavioral outcome.
 \item \textbf{Phase~I (Feb--Nov 2023).} A rider-heterogeneous binary logit with random mode slopes, heterogeneity in means, Mundlak user-mean corrections, and city~$+$~month-of-year fixed effects generates model-implied predictions for each outcome under a within-user TUB~$\to$~governed substitution.
 \item \textbf{Phase~II (Dec 2023).} A trip-level linear probability model with rider, city, and month-of-year fixed effects (identical FE block to Phase~I) tests those predictions causally, using cross-city variation in pre-ban TUB share as continuous treatment intensity (Bonferroni $\alpha_{\text{Bonf}} = 0.025$).
 \item \textbf{Composition check.} A within-user mode-switcher design verifies that the causal drop is a within-user reduction, not a compositional offset from aggressive TUB riders shifting into the governed pool.
\end{itemize}


This paper makes three contributions. First, we provide the first rider-heterogeneous random-parameter cross-sectional prediction of a micromobility speed-governance intervention that is matched on the same binary probability scale by a user-FE causal DiD, jointly on harsh acceleration and harsh deceleration, under a single user~$+$~city~$+$~month-of-year identification axis shared by Phase~I and Phase~II. Second, we estimate a jointly Bonferroni-significant reduction in both harsh-event outcomes (both harsh acceleration and harsh deceleration decline, rather than moving in opposite directions) on a 3-month pre-trend parallel-trends window that both outcomes satisfy. Third, we exploit a natural experiment, the December~2023 system-wide removal of Swing's ungoverned mode across 52 cities, to test the Phase~I cross-sectional predictions causally. Both predictions are confirmed in direction and order of magnitude, and the residual gap between prediction and causal estimate is accounted for by an explicit sample re-weighting decomposition between the mode-stratified Phase~I fit and the full Phase~II trip universe, not by within-user behavioral change.

Section~\ref{sec:Section2} develops the research hypotheses. Section~\ref{sec:EmpiricalStrategy} describes data and methods. Sections~\ref{sec:phase1_results}--\ref{sec:res_compensation} report the cross-sectional predictions, causal validation, and composition check. Sections~\ref{sec:discussion}--\ref{sec:conclusion} close.

\section{Related literature}
\label{sec:Section2}

\subsection{South Korean regulatory context.}

South Korea introduced formal regulations for personal mobility devices (PM) through amendments to the Road Traffic Act (\textit{Doro Gyotong-beop}) effective in May 2021. The amendments classified e-scooters as a subcategory of motorized bicycles, imposing a maximum speed limit of 25~km/h, a minimum rider age of 16, mandatory helmet use, and a prohibition against sidewalk riding. Operators of shared e-scooter services were required to implement firmware speed governors ensuring that vehicles do not exceed the 25~km/h threshold. In practice, enforcement of rider-side rules (helmets, sidewalk riding, age) has been limited, leaving the firmware governor as the dominant policy instrument and providing a well-defined setting for isolating its effect.

\subsection{E-scooter injury epidemiology and the speed--severity gradient.}

The first wave of academic e-scooter research was dominated by clinical and emergency-department studies that documented the scale and pattern of injuries associated with shared e-scooter adoption. \citet{Aizpuru_2019} reviewed a decade of motorized-scooter presentations in the U.S. National Electronic Injury Surveillance System and showed that injury volumes rose substantially with the arrival of dockless sharing services, with head injuries and fractures dominating. \citet{Sikka_2019} reported a case of pedestrian injury caused by an e-scooter rider on a sidewalk, illustrating that injury burden extends beyond riders themselves. \citet{Bloom_2021} found that standing e-scooter injuries in a major U.S. community imposed material trauma-center demand, with head trauma and upper-limb fractures the most common patterns. \citet{Nellamattathil_2020} confirmed similar patterns in a separate metropolitan area after widespread e-scooter adoption. More recently, \citet{Hoveidaei_2025} used a U.S. nationwide database to show that orthopedic fractures associated with e-scooter use rose steeply relative to skateboard and traditional scooter injuries between 2010 and 2022, while \citet{ArikanOzturk_2024} provided the first systematic police-report analysis of e-scooter accidents in T\"urkiye. \citet{Cittadini_2022} found clinically significant protective effects of helmet use against head-injury severity in Italian e-scooter crashes, reinforcing the role of crashworthiness in injury outcomes.

Speed is the most consistent severity determinant in this literature. \citet{Cicchino_2021} linked trip characteristics, including speed and trip purpose, to injury severity among hospitalized e-scooter riders, and \citet{Yang_2020} used news-report mining to show that collisions with motor vehicles and loss-of-control events, both tightly linked to speed, drive the most severe outcomes. These e-scooter-specific results are consistent with the broader power model of the speed--crash relationship developed and updated by \citet{Elvik_2013,Elvik_2019}, which shows that mean traffic speed enters accident and fatality counts with an elasticity well above one, and that the gradient is especially steep for vulnerable road users. \citet{Abdi_2025} extended this line by linking built-environment typologies to e-scooter crash severity using random-parameter models, showing that road geometry and land-use mix materially shape crash-severity outcomes beyond individual rider-level factors. The epidemiological literature motivates speed as the primary modifiable risk factor, but it does not itself evaluate the instruments (governors, geofencing, and nighttime restrictions) that jurisdictions have deployed to manage it.

\subsection{Empirical evaluations of speed governance and related restrictions.}

The empirical literature that directly evaluates micromobility speed and usage restrictions is still small, but it is growing and methodologically heterogeneous. \citet{Quddus_2025} provide a recent and systematic causal evaluation of a large-scale speed-limit reduction (the UK's 20~mph roll-out), quantifying speed and safety impacts and showing that modest speed reductions translate into non-negligible casualty declines, a direct motor-vehicle analogue of the behavioral response we test here at the e-scooter level. \citet{Haworth_2021} document how shared versus private e-scooter use evolved in Brisbane after regulatory changes and discuss the safety implications for a heterogeneous rider mix, anticipating the compositional dimension that our DiD must separate from behavioral compensation. \citet{Kazemzadeh_2025} provides complementary evidence on how shared-space rider perceptions respond to context, reinforcing that riders' subjective speed tolerance is neither fixed nor uniform across environments.

Two clinical cohort studies exploit operator-imposed restrictions in Finnish cities. \citet{Liukkonen_2023} study a nighttime-only speed-limit intervention in Tampere (15~km/h between 00:00 and 06:00) and find that the restriction is not associated with a reduction in e-scooter injury incidence. \citet{Pakarinen_2023} evaluate a broader package in Helsinki that combines a daytime 20~km/h cap, a 15~km/h nighttime limit, and weekend overnight bans, and report a roughly 50\% drop in injuries per ride. Both studies treat injury counts as the outcome and do not resolve within-trip behavioral responses. In particular, neither design can distinguish whether any change in injury counts is driven by fewer or shorter trips, by mechanical speed capping, or by rider-side behavioral adjustment, and the contrast between the two Finnish results is itself consistent with instrument-specific mechanical channels rather than a uniform rider response. Our study directly addresses this decomposition by measuring behavior at the trip level.

\citet{Asensio_2022} exploit a geofencing-based natural experiment to identify the effect of a local micromobility regulation on car displacement and congestion, demonstrating that digital, firmware-level policy levers can be evaluated with causal designs. Their outcomes, however, concern aggregate mode substitution rather than within-trip rider behavior. \citet{Cloud_2023} push the identification strategy further by using staggered rollout of shared e-scooter services across 93 cities in six European countries to provide the first multi-country causal estimate that e-scooter availability raises police-reported traffic accidents. Their design shows that a causal evaluation of micromobility regulation is feasible, but the outcome is accident counts rather than behavioral margins. \citet{Caggiani_2023} propose a geofencing methodology for speed-limit regulation in e-scooter systems, and \citet{Lawrence_2020} evaluate a 30~km/h speed limit trial in Melbourne with before--after data, documenting speed reductions and modest safety improvements. \citet{Kwon_2024} analyze e-scooter risk factors in Korea at different speed levels and road types, finding that the marginal risk contribution of speed is strongly road-type dependent. Finally, \citet{Yasmin_2022} show that crash-frequency and severity models that ignore the endogeneity of speed enforcement systematically mis-identify the safety effect of speed governance, a caveat that is directly relevant to our identification strategy.

\subsection{Naturalistic riding, kinematic metrics, and within-trip risk signals.}

Advances in naturalistic driving data have made it possible to characterize risk at the level of individual trips rather than crash counts. \citet{Ma_2021} used mobile-sensing data to derive surrogate safety measures for e-scooter riders and showed that within-trip kinematic signals vary systematically with infrastructure type. \citet{White_2023} conducted a naturalistic riding study of e-scooters and identified trip- and environment-level factors associated with elevated crash risk, while \citet{Tian_2024} combined naturalistic riding data with crash records to characterize e-scooter-specific riding behaviors and their hazard profile. \citet{Cho_2023} derived riding-risk precursors from a naturalistic study of 100 delivery motor scooters, and \citet{Gu_2025} extended this work by associating sensor-based riding behavior with road-traffic characteristics to identify high-risk operating conditions. \citet{Baby_2024} developed and validated an e-scooter Riding Behavior Questionnaire for Korean riders, providing one of the first psychometrically validated self-report instruments for this population.

Within the broader driver-telematics literature, \citet{Ziakopoulos_2024} uses smartphone telematics and imbalanced-learning methods to model harsh braking and harsh acceleration occurrence, showing that these events are useful proxies for crash risk and are shaped by both road and driver characteristics. Recent reviews that span crash safety and driving-behavior modeling \citep{Petraki_2026} treat harsh events, jerk, and speed variability as standard within-trip indicators of unsafe riding or driving. Trajectory-level generative models of GPS rider and driver behavior \citep{Choi_2021_TrajGAIL}, together with broader surveys of deep generative methods in transportation \citep{Choi_2025_GenerativeReview}, offer a complementary lens on this literature, since they reproduce full speed and position sequences rather than reducing riding behavior to summary indicators. This body of work provides the methodological foundation for the behavioral outcomes we use, but, critically for this study, it does not connect kinematic metrics to a specific speed-governance policy. Our paper is the first to measure changes in harsh events, speed variability, cruising fraction, and zero-speed fraction around a system-wide removal of an ungoverned e-scooter mode.

\subsection{Micromobility demand, usage patterns, and platform retention.}

A complementary strand of literature studies micromobility demand and usage patterns without focusing on injury outcomes. \citet{Weschke_2023} estimates the effect of public-transport station proximity on shared e-scooter demand, showing that e-scooters complement rather than substitute for transit. \citet{Liu_2022} and \citet{Su_2024} examine how built-environment features and socioeconomic characteristics shape the spatial distribution of e-scooter use, with \citet{Su_2024} documenting pronounced spatial inequities between shared e-scooters and docked bikeshare in Washington, DC. \citet{Aarhaug_2025} compare shared and personal e-scooter usage patterns and mode-replacement effects, finding systematic differences that are informative about the underlying user mix. \citet{Currie_2025} analyze shared e-scooter trip patterns and their links to public-transport service quality in an Australian setting.

These studies describe who uses e-scooters and where, and they quantify the complementarity between shared e-scooters and transit. What they do not examine is how \textit{trip-level riding behavior} responds to a policy intervention that changes the product itself, specifically the removal of an ungoverned mode. Our paper focuses on that behavioral margin, leaving demand and retention responses to future work on the same natural experiment.

\subsection{Positioning of this study and research hypotheses.}

Against this background, our empirical analysis tests the following hypothesis.

\begin{description}
\item[H1 (Joint harsh-event reduction).] Removal of the TUB mode causally reduces the trip-level probability of both harsh acceleration ($\mathbf{1}[\text{any harsh accel event}]$) and harsh deceleration ($\mathbf{1}[\text{any harsh decel event}]$), with both outcomes declining together rather than moving in opposite directions. This is the paper's primary behavioral hypothesis and is the target of the Phase~I$\to$Phase~II predict-then-validate design.
\end{description}

\noindent Because the firmware governor on STD and ECO physically prevents speeds above $25$~km/h, a speeding indicator ($\mathbf{1}[\max\text{speed}>25\text{ km/h}]$) is carried through the analysis as a manipulation check to confirm that the policy operates at the firmware level as intended, but it is not a behavioral hypothesis and is excluded from the primary outcome family (see Section~\ref{sec:outcomes} and Appendix~\ref{app:speeding_check}).

\section{Methods}
\label{sec:EmpiricalStrategy}

\subsection{Data and the December 2023 natural experiment}
\label{sec:data}

\subsubsection{The Swing platform.}
\begin{figure}[width=0.99\textwidth, pos = t]
 \centering
 \includegraphics[width=0.99\textwidth]{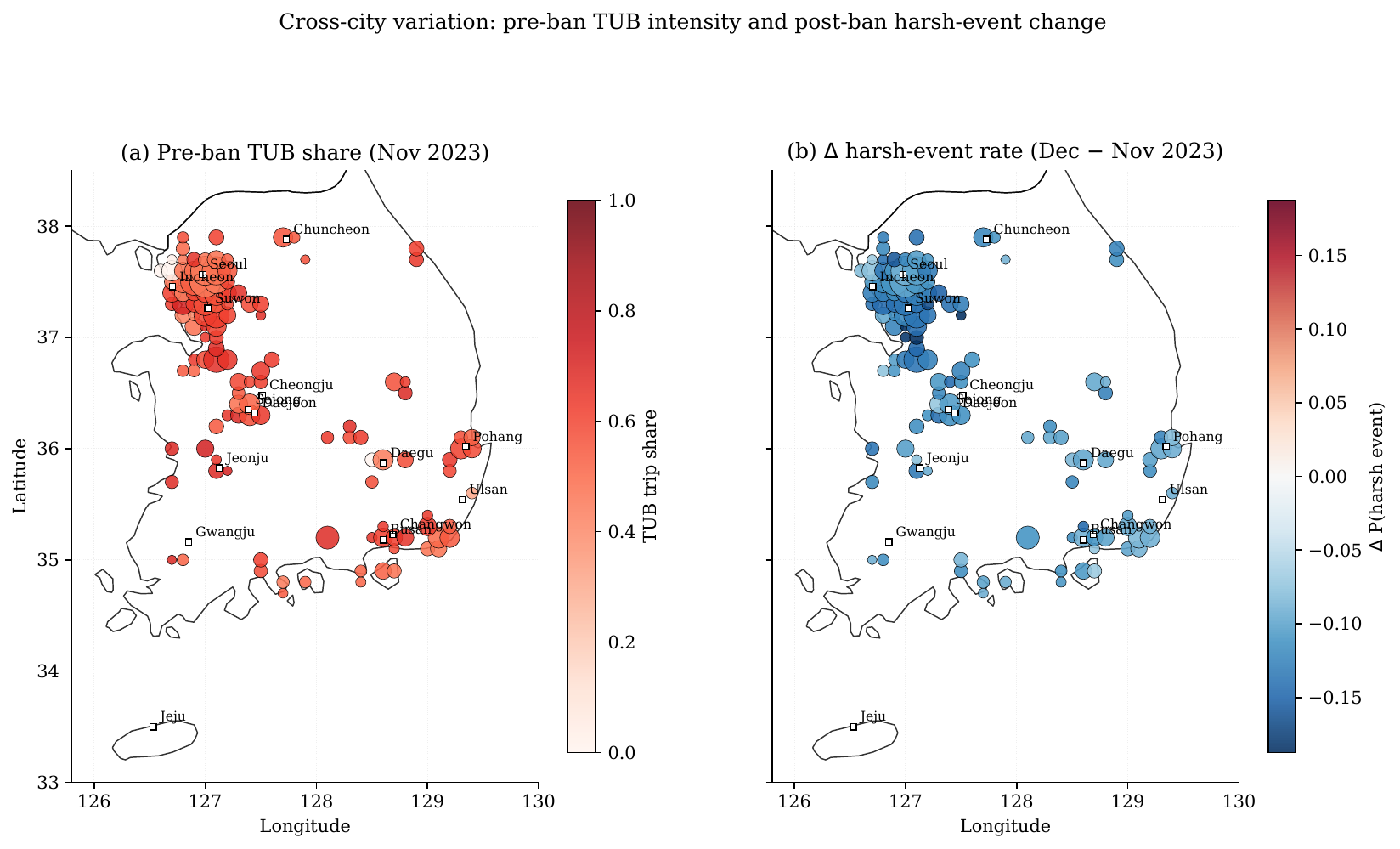}
 \caption{Cross-city variation in pre-ban TUB intensity and post-ban harsh-event reduction. \textbf{Left}, November~2023 TUB trip share by city. \textbf{Right}, change in trip-level harsh-event probability (Dec $-$ Nov~2023). Circle size is proportional to $\sqrt{N_{\text{trips}}}$.}
 \label{fig:swing_cities}
\end{figure}

Swing is the largest dockless shared e-scooter operator in South Korea by fleet size, accounting for roughly one-third of the approximately $300{,}000$ shared micromobility devices deployed nationwide as of 2023--2024 and operating approximately $90{,}000$--$100{,}000$ devices across its fleet of e-scooters, shared bicycles, and e-mopeds. Its behavioral data therefore provides broad coverage of the Korean shared-e-scooter population rather than a niche or operator-specific sample. During the study period (Feb--Nov~2023) Swing operated across 52 cities spanning the full geographic extent of South Korea, including all seven metropolitan cities, Sejong, provincial capitals, and mid-sized municipalities. This geographic diversity produces substantial cross-city variation in baseline TUB intensity (Figure~\ref{fig:swing_cities}, left), which Phase~II exploits as continuous treatment. The right panel previews the post-ban harsh-event reduction, which is larger in cities with higher pre-ban TUB share.

\begin{table}[width=0.99\textwidth, pos = !hb]
\centering
\caption{Dataset overview.}
\label{tab:dataset_overview}
\begin{tabular}{ll}
\toprule
\textbf{Attribute} & \textbf{Value} \\
\midrule
Total trips & 19,458,758 \\
Unique users & 1,001,459 \\
Cities & 52 \\
Study period & February--November 2023 \\
TUB (turbo) trips & 10,181,605 (52.3\%) \\
STD (standard) trips & 8,307,655 (42.7\%) \\
ECO (economy) trips & 969,498 (5.0\%) \\
GPS sampling interval & $\sim$10 seconds \\
\bottomrule
\end{tabular}
\end{table}

\subsubsection{Speed modes and same-hardware design.}

Swing offered three speed modes on identical vehicle hardware (same motor, battery, chassis, and brakes), distinguished only by a firmware-level speed cap. These modes are \textit{Turbo} (TUB, ungoverned, exceeding 25~km/h), \textit{Standard} (STD, $\approx$25~km/h), and \textit{Economy} (ECO, $\approx$20~km/h). Riders selected a mode at the start of each trip. Because nothing but the governor differs across modes, any behavioral difference is attributable to speed governance rather than to vehicle characteristics, providing a quasi-experimental setting for isolating governance effects.

\subsubsection{Dataset and variable construction.}

We observe $19{,}458{,}758$ complete trips from $1{,}001{,}459$ unique riders across the 52 operating cities over Feb--Nov~2023 (Table~\ref{tab:dataset_overview}). Each trip record contains a rider identifier, a mode flag (TUB, STD, or ECO), start and end GPS coordinates, a start timestamp, total distance and travel time, and a sequence of on-board speed measurements sampled at approximately $10$-second intervals from the e-scooter's sensor telemetry. Revealed preference favored the ungoverned mode (TUB $52.3\%$, STD $42.7\%$, ECO $5.0\%$).

\begin{figure}[width=0.99\textwidth, pos = !t]
\centering
\includegraphics[width=0.99\textwidth]{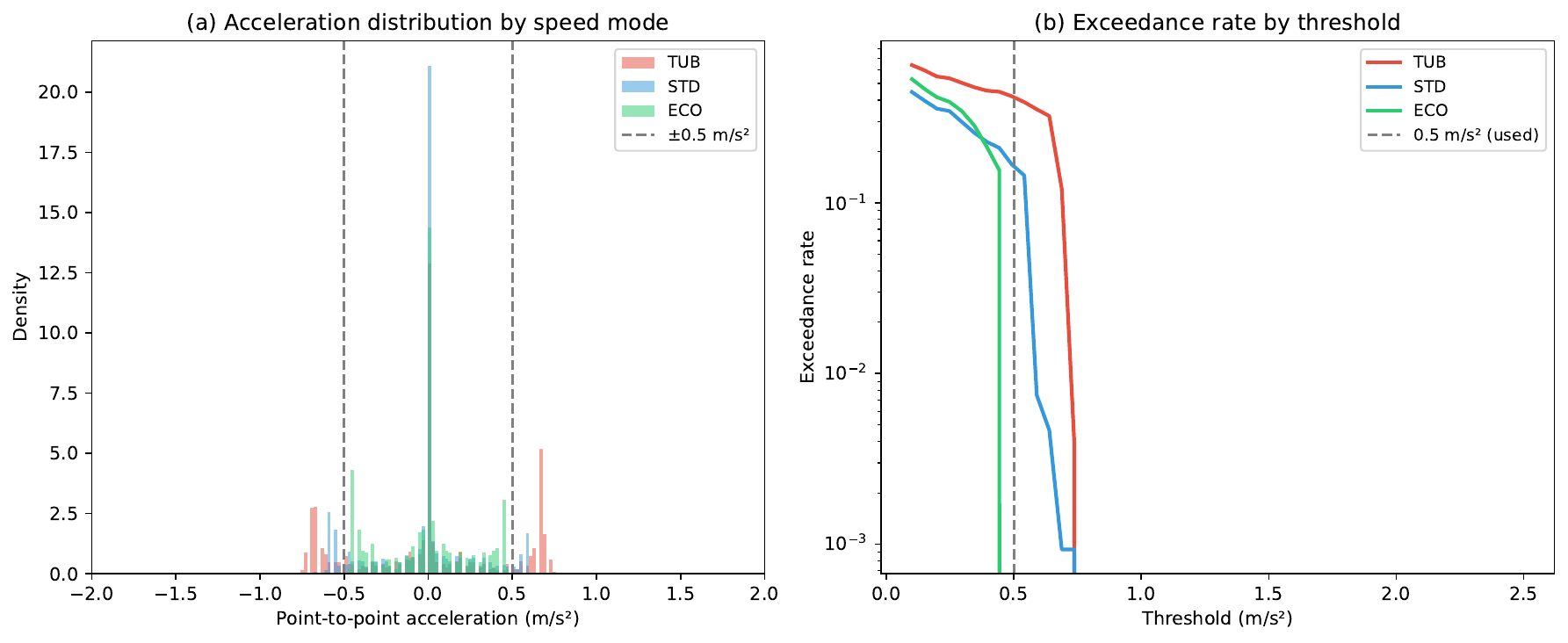}
\caption{Acceleration distribution and harsh-event threshold calibration. (a)~Point-to-point acceleration distribution by speed mode with the $\pm 0.5$~m/s$^2$ threshold (dashed lines). (b)~Exceedance rate as a function of threshold.}
\label{fig:accel_dist}
\end{figure}

\begin{table}[width=0.99\textwidth, pos = !hb]
\centering
\caption{Trip-level means (SD) by speed mode, Feb--Nov~2023.}
\label{tab:descriptive}
\small
\begin{tabular}{llrrr}
\toprule
\textbf{Category} & \textbf{Outcome} & \textbf{TUB} & \textbf{STD} & \textbf{ECO} \\
\midrule
\textit{Speed} & Mean speed (km/h) & 15.730 (4.432) & 13.427 (3.874) & 10.672 (3.329) \\
\addlinespace
\multirow{3}{*}{\textit{Safety}}
& Harsh acceleration (count/trip) & 0.480 (0.840) & 0.233 (0.564) & 0.020 (0.204) \\
& Harsh deceleration (count/trip) & 0.675 (0.921) & 0.332 (0.635) & 0.020 (0.205) \\
& Speed CV & 0.564 (0.301) & 0.558 (0.301) & 0.564 (0.314) \\
\addlinespace
\multirow{2}{*}{\textit{Dynamics}}
& Cruise fraction & 0.259 (0.202) & 0.320 (0.249) & 0.416 (0.302) \\
& Zero-speed fraction & 0.155 (0.145) & 0.155 (0.148) & 0.167 (0.155) \\
\midrule
& $N$ trips & 10{,}181{,}605 & 8{,}307{,}655 & 969{,}498 \\
& Mode share & 52.3\% & 42.7\% & 5.0\% \\
\bottomrule
\end{tabular}
\end{table}

From the speed sequence of each trip we construct all behavioral variables used in the analysis, including maximum and mean speed, the within-trip speed coefficient of variation, cruise fraction (share of waypoints within $\pm 3$~km/h of the trip mean), zero-speed fraction, and the two primary binary harsh-event outcomes (Section~\ref{sec:outcomes}). Point-to-point acceleration at lag $\Delta t = 10$~s is computed from consecutive speed waypoints, and the harsh-event indicators are trip-level $\mathbf{1}[\cdot]$ of whether any such consecutive-pair acceleration exceeds the $\pm 0.5$~m/s$^2$ threshold (Figure~\ref{fig:accel_dist}). Because every outcome is derived from the same speed sequence, any measurement error in the underlying sensor is common across outcomes and is absorbed by rider and city fixed effects in the Phase~II DiD. Trip-level context covariates ($\mathbf{1}[\text{night}]$ for a start between 22:00--06:00, $\mathbf{1}[\text{weekend}]$, $\mathbf{1}[\text{urban}]$ for a trip originating in one of the nine metropolitan cities, $\mathbf{1}[\text{same route}]$ for a start in a $500$~m grid cell previously visited by the same rider, and $\log(1 + \text{consecutive riding days})$) are constructed from each rider's ordered trip history. We apply minimal trip-level restrictions (duration $\geq 30$~seconds and at least three valid speed waypoints) to remove lock/unlock artifacts and short trial rides.

Table~\ref{tab:descriptive} reports trip-level means (and standard deviations) of the main behavioral variables by speed mode. Harsh events show an order-of-magnitude gradient across modes. TUB generates $0.48$ harsh-accel events per trip versus $0.02$ for ECO, a $25$-fold difference, and the analogous gradient on harsh deceleration is $0.68 \to 0.34 \to 0.02$. Mean speed follows the same monotonic pattern ($15.7 \to 13.4 \to 10.6$~km/h). Speed CV is nearly flat across modes ($0.561$--$0.569$), which is why the Phase~I/II analysis retains the two harsh-event outcomes as the primary behavioral margins and leaves CV-style dynamics variables to the composition check of Section~\ref{sec:res_compensation}.

\subsubsection{The December 2023 TUB ban.}

In December 2023, Swing removed the ungoverned mode system-wide in response to regulatory guidance, reducing TUB's trip share from 54.6\% to 1.4\% within a single month. The residual 1.4\% reflects trips initiated during the brief transition window at the start of December. This removal constitutes a simultaneous, system-wide shock across all 52 cities, avoiding the staggered-adoption concerns highlighted by \citet{GoodmanBacon_2021}. Because the shock is driven by a national regulatory directive rather than by city-level safety conditions or demand patterns, it is plausibly exogenous to the behavioral outcomes we study.

\subsection{Identification strategy}
\label{sec:Identification}

We identify the causal effect of the December~2023 TUB ban by comparing the pre-to-post change in trip-level harsh-event probabilities across cities that differed in their pre-ban TUB intensity. This is a difference-in-differences design with two differences. The first is temporal (December versus the pre-ban months), and the second is cross-sectional (cities with higher versus lower pre-ban TUB share). Because the ban was enacted simultaneously across all 52 Swing cities as a nationwide regulatory directive, there is no never-treated control city. Identification instead exploits the \textit{continuous cross-sectional variation} in pre-ban TUB share, which ranged from about $0.003$ to $0.703$. Under the parallel-trends assumption that cities with higher and lower pre-ban TUB share would have followed the same outcome trend absent the ban, the slope of the pre-to-post change in outcome against pre-ban TUB share identifies the causal effect of the policy on a per-unit-TUB-share basis. We test this assumption directly on a 3-month pre-ban window in Section~\ref{sec:phase2_results}.

This design maps onto the continuous-treatment difference-in-differences framework of \citet{Callaway_2024}, which extends DiD identification to continuous treatment intensities under a simultaneous shock. The two-way fixed-effects implementation is equivalent to the two-way Mundlak regression of \citet{Wooldridge_2025}, which is the perspective we adopt for the trip-level linear probability model below. The same-hardware design of Section~\ref{sec:data} rules out between-mode vehicle heterogeneity as a confound, and all trips are observed on a single platform, eliminating between-operator confounding. The baseline specification has user, city, and month-of-year fixed effects on each of the two outcomes (harsh acceleration and harsh deceleration), with Bonferroni correction across the two-outcome family ($\alpha_{\text{Bonf}} = 0.05/2 = 0.025$).

\subsection{Outcome definitions}
\label{sec:outcomes}

For each trip $i$ with onboard speed vector $\mathbf{v}_i = (v_{i1}, \ldots, v_{iT_i})$ recorded at approximately 10-second intervals, we compute two primary trip-level \textit{binary} safety outcomes and one firmware-validation indicator. Each outcome is a $\{0,1\}$ indicator at the trip level, so the same variable can serve as the dependent variable for both the rider-heterogeneous logit in Phase~I and the user-FE difference-in-differences in Phase~II.

\textit{Primary Outcome 1, Harsh acceleration.} $\text{HarshAccel}_i = \mathbf{1}[\exists\,t \ \text{with} \ a_{it} > 0.5~\text{m/s}^2]$, where $a_{it} = (v_{i,t+1} - v_{it}) / (\Delta t \cdot 3.6)$ is the point-to-point acceleration (km/h converted to m/s) at $\Delta t \approx 10$~s sampling. The $0.5$~m/s$^2$ threshold is calibrated to the $97.5$th percentile of the absolute acceleration distribution across $6.3$ million speed-point pairs in a $200{,}000$-trip sample (Figure~\ref{fig:accel_dist}) and is not chosen to fit the result. A conventional $2$~m/s$^2$ automotive threshold \citep{Ziakopoulos_2024} yields near-zero events at the $10$-second sampling interval ($<$0.01\% exceedance for all modes), precluding meaningful analysis, so the micromobility-calibrated threshold is the only workable choice for GPS-based e-scooter telemetry. The binary random-parameters logit specification for a trip-level safety outcome follows recent e-scooter and cyclist crash work \citep{Agheli_Cyclist_2025,Scarano_2025}, and the trip-level ``any harsh event'' indicator itself is grounded in the naturalistic-driving safety-event literature \citep{Bagdadi_2013,Ziakopoulos_2024}.

\textit{Primary Outcome 2, Harsh deceleration.} $\text{HarshDecel}_i = \mathbf{1}[\exists\,t \ \text{with} \ a_{it} < -0.5~\text{m/s}^2]$, symmetric to Primary Outcome~1.

\textit{Firmware-validation check, Speeding.} $\text{Speeding}_i = \mathbf{1}[\max_t v_{it} > 25~\text{km/h}]$, where $25$~km/h is the statutory e-scooter ceiling in Korea. This indicator is \textit{not} treated as a behavioral outcome. Because the firmware governor on STD and ECO physically prevents exceedance, any Phase~II coefficient on speeding is a hardware confirmation of the treatment rather than a causal behavioral response, and Bonferroni correction is therefore applied to the two-outcome harsh-event family only.

The two primary outcomes are restricted to the harsh-event margins because these are the direct behavioral channels a firmware speed cap can plausibly affect. Three additional riding-style variables (speed coefficient of variation, or CV, cruising fraction as time share within $\pm 3$~km/h of the trip mean, and zero-speed fraction) are \textit{not} used as primary outcomes, since they reflect trip context such as traffic, geography, and trip purpose rather than kinematic risk per se. They are nonetheless useful as a diagnostic of whether within-user behavior shifts on \textit{any} behavioral margin after losing TUB. If riders compensated for the lost top speed by driving more aggressively on the governed modes, this within-user switcher comparison on the two harsh-event margins should detect it; the remaining kinematic variables (mean speed, CV, cruise, zero-speed fraction) are used only for descriptive context and not as compensation-check outcomes. We use them in the composition check of Section~\ref{sec:res_compensation} to rule out precisely this type of compositional offsetting.

\subsection{Empirical model}
\label{sec:meth_models}

Both Phase~I and Phase~II are estimated at the trip level with user, city, and month-of-year fixed effects. Phase~I allows mode slopes to vary across riders (random parameters), while Phase~II identifies a single causal effect of the December~2023 ban (difference-in-differences). The shared trip universe and fixed-effect block let the Phase~I prediction and Phase~II causal estimate be compared directly on each binary outcome.

\subsubsection{Phase I: rider-heterogeneous random-parameters binary logit}
Phase~I fits a trip-level random parameters binary logit separately for each of the two primary harsh-event outcomes (and, in parallel, for the speeding firmware-validation check of Appendix~\ref{app:speeding_check}). Following standard practice for random-parameters models of crash and behavioral outcomes \citep{Mannering_2016,Ahmed_2023,Alnawmasi_2025,Agheli_2025,Agheli_Cyclist_2025}, the log-odds of each binary outcome is modeled as
\begin{equation}
 \text{logit}\,\Pr\!\big(Y_{it}^{(j)} = 1\big) = \beta_{1i}\,\text{TUB}_{it} + \beta_{2i}\,\text{ECO}_{it} + \boldsymbol{\gamma}^{\prime} Z_i + \boldsymbol{\delta}^{\prime} W_{it},
 \label{eq:rplogit}
\end{equation}
where the rider-specific mode slopes $\beta_{ki}$ are drawn from a normal whose mean is allowed to shift with the rider's cumulative pre-trip experience $z_i = \log(1 + \text{cum.\ trips}_{it})$,
\begin{equation}
 \beta_{ki} \sim \mathcal{N}\!\left(\mu_k + \Gamma_k\,z_i,\ \sigma_k^2\right), \qquad k \in \{\text{TUB}, \text{ECO}\}.
 \label{eq:rplogit_hetmean}
\end{equation}
$\sigma_k$ measures how widely the mode effect varies across riders, and a large and significant $\sigma_k$ is direct evidence that a homogeneous-slope model misses genuine rider heterogeneity. The parameter $\Gamma_k$ lets the mean mode effect grow or shrink with experience, following the ``heterogeneity in means'' device of \citet{Mannering_2016}. We implement this device asymmetrically across the two random mode slopes: the TUB slope shifts with three rider-level moderators (cumulative riding experience, night-trip context, and same-route familiarity) while the ECO slope shifts only with cumulative experience. The asymmetry is substantive rather than incidental. TUB is the policy-relevant treatment mode whose behavioral margin we predict and the December~2023 ban removes, so identification of \textit{who} is more or less harsh under TUB carries the policy content; ECO and STD operate under the same firmware speed cap and serve as governed-mode baseline categories, on which we model only the standard within-mode learning gradient via experience. Concentrating the heterogeneity-in-means parameters on the TUB slope also keeps the simulated-MLE estimator stable, since the ECO mass share in the stratified pre-ban fit sample is small relative to TUB and additional ECO interactions would inflate standard errors without improving identification on the policy-relevant margin. The trip-level control vector $W_{it}$ collects five exogenous indicators, namely $\mathbf{1}[\text{night}]$ (trip started between 22:00 and 06:00), $\mathbf{1}[\text{weekend}]$, $\mathbf{1}[\text{same route}]$ (trip started from the same 500\,m grid cell as one of the rider's previous trips, a proxy for route familiarity), and $\log(1 + \text{consec.\ riding days})$ (the number of consecutive days the rider has used the platform, a recency measure). The urban indicator (origin in one of Korea's nine metropolitan cities), which is invariant within city, is omitted from the regressor list in both phases because it is collinear with the city fixed effects. $Z_i$ contains rider-level means of these five covariates plus $\{\text{TUB}, \text{ECO}\}$, following the Mundlak correction \citep{Mundlak_1978} so that the rider-specific slopes are identified from within-user variation rather than from differences between rider types. Each logit is fit by simulated maximum likelihood with $200$ Halton draws on a stratified pre-ban (Feb--Nov~2023) sample, with the rider panel imposed so that all of a rider's trips share the same random-parameter draw.

The Phase~I logit answers a simple what-if question for each harsh-event outcome. Had every pre-ban TUB trip been ridden in the governed mode instead, by how much would the average event probability have fallen? The answer is a within-user prediction, not a causal effect. It serves as a numerical target that the Phase~II DiD then tests directly.

\subsubsection{Phase II: difference-in-differences (DiD) specification}
Phase~II estimates the causal effect of the December~2023 TUB ban on each of the two primary harsh-event binary outcomes using a trip-level linear probability model with user, city, and month-of-year fixed effects (the same FE structure as Phase~I) on the same pooled trip universe used to construct the Phase~I outcomes.\footnote{In the linear Phase~II specification, user fixed effects are absorbed as literal user dummies via \texttt{pyfixest}. In the nonlinear Phase~I logit, the same user fixed effects are implemented via the \citet{Mundlak_1978} within-transformation (user means of all time-varying regressors plus a user-panel structure in which all of a user's trips share the same random-parameter draw), the standard approach in the AMAR random-parameter literature for avoiding the Neyman--Scott incidental-parameter problem in nonlinear MLE \citep{Mannering_2016}. The two implementations identify the same within-user variation.} Because the ban was simultaneous across all $52$ cities, we exploit cross-city variation in pre-ban TUB intensity as a continuous treatment, specified as
\begin{equation}
 Y_{it}^{(j)} = \alpha_i + \eta_{c(i)} + \gamma_{m(t)} + \beta^{(j)} \cdot \big(\text{Post}_t \times \text{TUB}_{c(i)}^{\text{Nov}}\big) + \varepsilon_{it},
 \label{eq:did}
\end{equation}
where $Y_{it}^{(j)}$ is trip $i$'s binary value on outcome $j$, $\alpha_i$ is a user fixed effect (absorbing all time-invariant user traits), $\eta_{c(i)}$ is a city fixed effect (absorbing city-level baselines), $\gamma_{m(t)}$ is a month-of-year fixed effect (absorbing national monthly time variation), $\text{Post}_t = \mathbf{1}[t = \text{Dec 2023}]$, and $\text{TUB}_{c(i)}^{\text{Nov}}$ is the November 2023 TUB trip share in the user's city, serving as a continuous treatment intensity. The parameter $\beta^{(j)}$ is the causal effect of governor imposition on outcome $j$, identified under the assumption that cities with higher and lower pre-ban TUB shares would have followed parallel trends absent the ban. Standard errors are clustered by city ($52$ clusters) via \texttt{pyfixest}'s CRV1 implementation \citep{fischer2024pyfixest}. Bonferroni correction is applied with $\alpha_{\text{Bonf}} = 0.05 / 2 = 0.025$ across the two primary harsh-event outcomes. The use of cross-sectional pre-treatment exposure intensity ($\text{TUB}_{c(i)}^{\text{Nov}}$) interacted with a single national post-treatment indicator places this design in the long-established continuous-treatment DiD tradition initiated by \citet{Card_1992}, with identification, weighting, and event-study aggregation properties under the modern continuous-treatment framework formalized by \citet{Callaway_2024} and \citet{CallawayES_2024}.

\begin{figure}[width=0.99\textwidth, pos = !ht]
\centering
\includegraphics[width=0.85\textwidth]{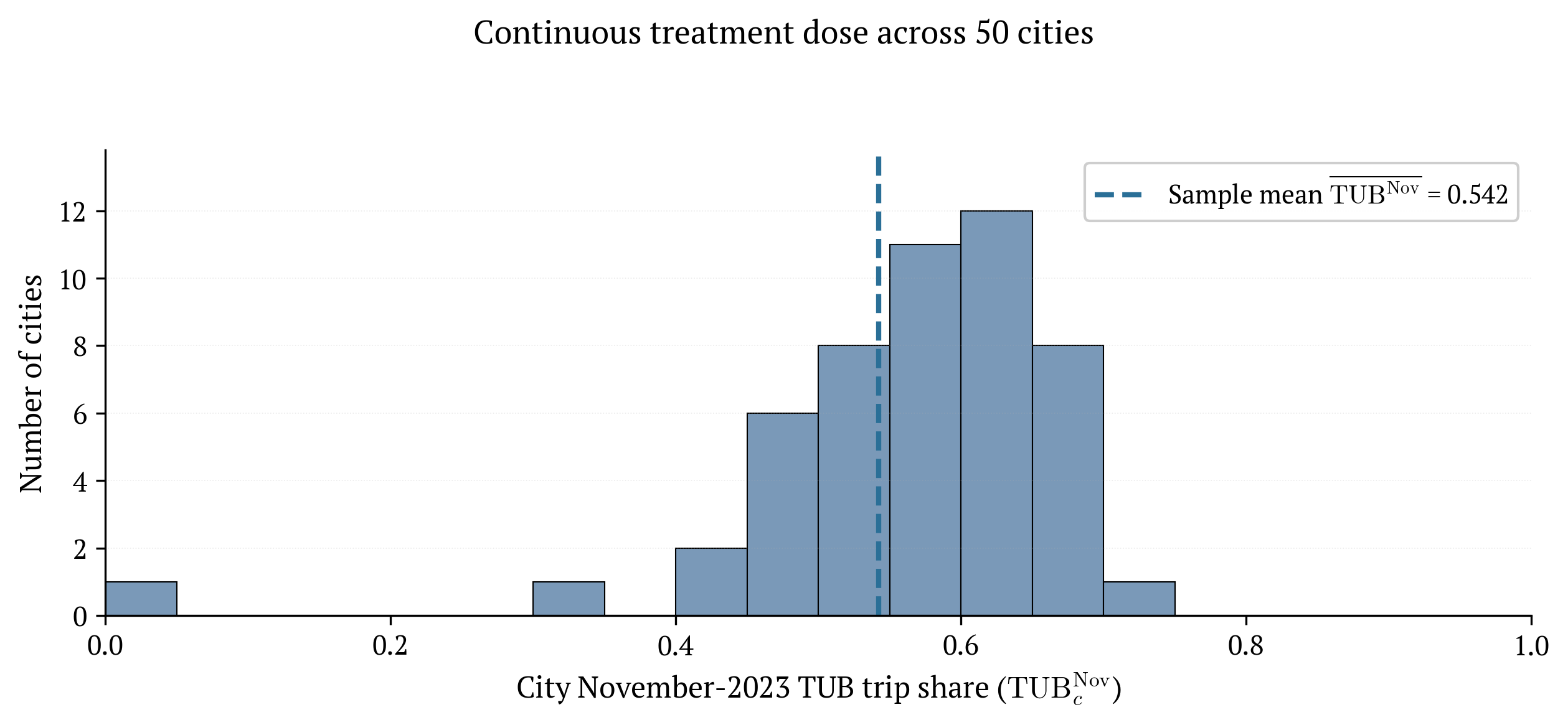}
 \caption{Distribution of the continuous treatment dose $\text{TUB}^{\text{Nov}}_{c}$ across 50 of the 52 study cities. Dashed line
   marks the sample-mean dose $\overline{\text{TUB}^{\text{Nov}}} = 0.542$. Two cities (Jeju and Mokpo) are excluded because they   
  had no recorded trips in November 2023.}  
\label{fig:tub_share_hist}
\end{figure}

To fix intuition for the continuous-dose design, $\text{TUB}^{\text{Nov}}_{c}$ is not a treated/control indicator but a cross-city intensity measure that takes values between $0.003$ (Gimpo, where TUB was essentially absent pre-ban) and $0.703$ (Ansan, where TUB dominated the pre-ban mode mix), with a sample-mean treatment dose $\overline{\text{TUB}^{\text{Nov}}} = 0.542$ used below for the avg-city percentage-point scaling. Figure~\ref{fig:tub_share_hist} plots the full distribution across the $52$-city cross-section. Because no city is untreated in the binary sense and treatment intensity varies sharply across cities, the binary ``treated vs control'' DiD is not available here, and the continuous dose captures precisely how much of each city's pre-ban trip mix the December ban removed. The coefficient $\beta^{(j)}$ on the interaction therefore reads as the change in the trip-level probability of outcome $j$ per one-unit increase in the city's pre-ban TUB share, and its avg-city percentage-point conversion $\beta^{(j)} \times \overline{\text{TUB}^{\text{Nov}}} \times 100$ reports the implied change for a city at the sample-mean dose.

The specification keeps the structure minimal, with no user-month or city-month aggregation and no city-specific trend regressors. User fixed effects absorb all time-invariant user traits, city fixed effects absorb city-level baselines, and month-of-year fixed effects absorb nationwide monthly time dynamics. The causal coefficient $\beta^{(j)}$ is identified from trip-level deviations of each user's December outcome relative to their own pre-ban baseline, conditional on city and monthly national dynamics and weighted by the trip's city-level treatment intensity. An event-study variant replaces $\text{Post}_t$ with month-by-month interactions ($\sum_{m \neq \text{Nov 2023}} \text{TUB}_{c(i)}^{\text{Nov}} \cdot \mathbf{1}[t = m]$) to visualize parallel trends. The pooled and event-study regressions are fit on a $2$-million-trip rider-aware sub-sample of the full Feb--Dec 2023 trip universe, following the convention in existing micromobility user-FE analyses that keeps high-dimensional fixed-effect absorption tractable.

\subsection{Composition check}
\label{sec:meth_tests}

Aggregate DiD effects may reflect composition shifts (former TUB users entering the STD/ECO pool) rather than a within-user reduction. To verify that the Phase~II estimates are not a compositional artifact, we compare 24{,}357 ``mode-switchers'' (TUB users in Oct--Nov 2023 who rode STD/ECO in Dec 2023) against 79{,}700 ``never-TUB'' controls. For each user, we compute pre- (Oct--Nov) and post-period (Dec) outcome means restricted to STD/ECO trips, then estimate a user-level DiD:
\begin{equation}
 \text{DiD}_{\text{comp}}^{(j)} = \overline{\Delta Y}_{\text{switcher}}^{(j)} - \overline{\Delta Y}_{\text{never-TUB}}^{(j)}
 \label{eq:compensation}
\end{equation}
Significance is assessed via Welch's $t$-test, and effect sizes are reported as Cohen's $d$. If the city-level DiD shows significant effects but the within-user switcher DiD yields negligible Cohen's $d$ values ($|d| < 0.2$), the aggregate change is attributable to composition rather than masking a within-user offset.

\section{Results}
\label{sec:results}

\subsection{Phase I: cross-sectional predictions}
\label{sec:phase1_results}

Phase~I fits a rider-heterogeneous random parameters binary logit (Equation~\ref{eq:rplogit}) on each of the two primary harsh-event outcomes defined in Section~\ref{sec:outcomes}, namely \textit{harsh acceleration} ($\mathbf{1}[\text{any harsh accel event on the trip}]$) and \textit{harsh deceleration}. Each Phase~I estimate and each Phase~I prediction below maps one-to-one onto a Phase~II causal test in Section~\ref{sec:phase2_results}. The parallel firmware-validation fit on speeding is reported in Appendix~\ref{app:speeding_check}.

\subsubsection{Rider-heterogeneous binary model.}
Each column in Table~\ref{tab:rp_rich_profile} is a separate fit on a stratified pre-ban sample. The specification and FE block are as described in Section~\ref{sec:meth_models}. Two findings matter. First, \textit{the rider-level spread of the TUB mode effect is positive and highly significant for harsh acceleration} ($\sigma_{\text{TUB}} = +0.761$, $p < 0.001$), with harsh deceleration showing marginal heterogeneity ($+0.284$, $p = 0.070$). A homogeneous logit misses this dispersion entirely. Second, \textit{the TUB $\times$ experience interaction is null on both harsh-event outcomes} ($-0.020$ and $-0.030$, both ns), meaning that experienced TUB riders do \textit{not} generate disproportionately more harsh events on TUB as they accumulate exposure. The harsh-event dimension is thus a near-constant function of experience within mode.

\begin{table}[width=0.99\textwidth, pos = !b]
\centering
\caption{Phase~I rider-heterogeneous random parameters binary logit for the two harsh-event outcomes. Stratified pre-ban sample (Feb--Nov 2023) with random mode slopes, heterogeneity in means, Mundlak controls, and city $+$ month-of-year FE. Firmware-validation speeding fit in Appendix Table~\ref{tab:rp_speeding_check}. $^{*}\,p<0.05$, $^{**}\,p<0.01$, $^{***}\,p<0.001$. Standard errors in parentheses.}
\label{tab:rp_rich_profile}
\small
\begin{tabular}{lcc}
\toprule
 & \textbf{Harsh accel} & \textbf{Harsh decel} \\
\midrule
\multicolumn{3}{l}{\textbf{Random mode slopes (mean and spread)}} \\
\quad TUB (mean)         & $+0.599^{***}$ & $+0.734^{***}$ \\
                         & $(0.096)$      & $(0.083)$      \\
\quad ECO (mean)         & $-2.609^{***}$ & $-2.442^{***}$ \\
                         & $(0.497)$      & $(0.329)$      \\
\quad $\sigma_{\text{TUB}}$ & $+0.761^{***}$ & $+0.284$ \\
                            & $(0.099)$      & $(0.157)$      \\
\quad $\sigma_{\text{ECO}}$ & $+1.171^{*}$   & $+0.711$ \\
                            & $(0.473)$      & $(0.474)$      \\
\addlinespace
\multicolumn{3}{l}{\textbf{Heterogeneity in means (interactions)}} \\
\quad TUB $\times\,\log$ experience & $-0.020$       & $-0.030$      \\
                                     & $(0.024)$      & $(0.020)$     \\
\quad TUB $\times$ night             & $-0.016$       & $-0.090$      \\
                                     & $(0.065)$      & $(0.057)$     \\
\quad TUB $\times$ same route        & $-0.093$       & $-0.046$      \\
                                     & $(0.079)$      & $(0.069)$     \\
\quad ECO $\times\,\log$ experience  & $-0.084$       & $-0.149^{*}$  \\
                                     & $(0.056)$      & $(0.059)$     \\
\addlinespace
\midrule
\multicolumn{3}{l}{\textbf{Specification}} \\
\quad Unit of analysis & \multicolumn{2}{c}{Trip (one observation per trip)} \\
\quad User fixed effects 
& Yes & Yes \\
\quad City fixed effects     & Yes & Yes \\
\quad Month-of-year fixed effects  & Yes & Yes \\
\quad Controls: night, weekend, same route, consec.\ days, log experience & Yes & Yes \\
\midrule
\multicolumn{3}{l}{\textbf{Model-implied prediction (TUB $\rightarrow$ governed)}} \\
\quad Predicted $\Delta$ (pp)        & $-2.81$  & $-4.24$  \\
\quad Pr(outcome) as-is              & $0.1668$ & $0.2481$ \\
\quad Pr(outcome) TUB removed        & $0.1387$ & $0.2057$ \\
\midrule
\multicolumn{3}{l}{\textbf{Model fit}} \\
\quad Log-likelihood            & $-13{,}659$ & $-15{,}630$ \\
\quad McFadden pseudo-$R^{2}$   & $0.446$     & $0.367$     \\
\quad AIC                       & $27{,}474$  & $31{,}416$  \\
\quad BIC                       & $28{,}136$  & $32{,}077$  \\
\quad Estimated parameters $k$  & $78$        & $78$        \\
\quad Halton draws              & $200$       & $200$       \\
\midrule
\multicolumn{3}{l}{\textbf{Sample}} \\
\quad Trips          & $35{,}602$ & $35{,}602$ \\
\quad Riders         & $25{,}000$ & $25{,}000$ \\
\quad Outcome mean   & $0.1727$   & $0.2466$   \\
\bottomrule
\end{tabular}
\end{table}

\subsubsection{Temporal stability.}
Following \citet{Wang_2025} and \citet{Sadeghi_2024}, we split the pre-ban window into Feb--Jun and Jul--Nov halves and re-estimate the full Phase~I specification on each. The rider-heterogeneity parameter $\sigma_{\text{TUB}}$ is statistically indistinguishable across halves ($0.783 \to 0.785$, $Z_{\text{stab}} = -0.06$, $p = 0.96$), so the heterogeneity reported in Table~\ref{tab:rp_rich_profile} is not an artifact of any specific sub-period. The \textit{mean} TUB effect grows in magnitude from $+0.43$ to $+0.70$, directionally consistent with a strengthening safety case for the December ban rather than a specification breakdown. Full parameter-by-parameter stability test in Appendix Table~\ref{tab:temporal_stability}.

\subsubsection{Cross-sectional pattern under the TUB ban.}
The fitted logit lets us summarize the cross-sectional rider--mode association compactly. If every TUB trip in the pre-ban sample were within-user relabeled as a governed-mode trip and all other covariates were held fixed, what would the sample-average outcome probability be? The bottom panel of Table~\ref{tab:rp_rich_profile} reports these model-implied differences, with \textbf{$-2.81$~pp} for harsh acceleration ($0.1668 \to 0.1387$) and \textbf{$-4.24$~pp} for harsh deceleration ($0.2481 \to 0.2057$). These numbers describe the harsh-event differential between the same rider's TUB and governed trips in the pre-ban data. They are a cross-sectional pattern, not a causal counterfactual, because the rider-heterogeneous logit conditions only on observed covariates and user fixed effects rather than on a policy shock. Whether removing TUB actually moves the harsh-event probability in the same direction is a separate question that requires a policy-induced source of variation, which Section~\ref{sec:phase2_results} supplies via the December 2023 ban.

\begin{figure}[width=0.99\textwidth, pos = t]
\centering
\includegraphics[width=0.99\textwidth]{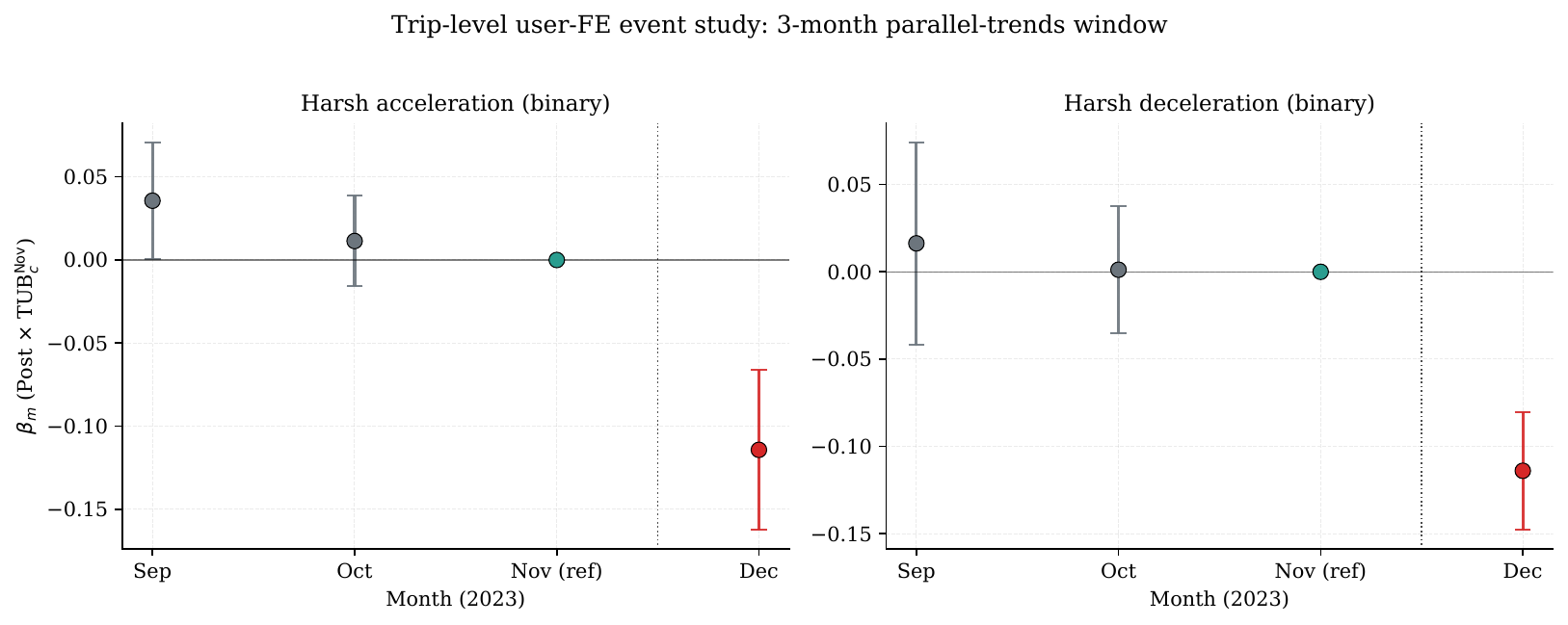}
\caption{Trip-level user-FE parallel-trends event study (Sep--Nov~2023 pre-ban, Nov as reference) for the two harsh-event outcomes, with $95\%$ city-clustered confidence intervals. Green marks the reference month, grey the pre-treatment months, and red the post-treatment month (Dec~2023). Vertical dotted line marks the policy break.}
\label{fig:event_study_user_fe}
\end{figure}

\subsection{Phase II: causal validation via the December 2023 ban}
\label{sec:phase2_results}

Whether eliminating TUB \textit{causes} the harsh-event probabilities to fall is a separate question from the cross-sectional association documented above. The December~2023 ban allows us to answer it using the trip-level linear probability model of Equation~\ref{eq:did}.

\subsubsection{Parallel trends and identification.}
The Phase~II DiD assumes that cities with higher and lower pre-ban TUB shares would have followed parallel trends absent the ban. We test this on the 3-month pre-ban window (Sep--Nov 2023, Nov as reference) using the trip-level user-FE event study in Figure~\ref{fig:event_study_user_fe}. \textbf{Both harsh-event outcomes satisfy parallel trends.} For harsh deceleration, neither pre-ban coefficient is distinguishable from zero. For harsh acceleration, neither is significant under Bonferroni correction ($\alpha_{\text{Bonf}} = 0.05/4 = 0.0125$, reflecting $2$ outcomes $\times$ $2$ pre-ban months); the September coefficient is marginal at the uncorrected $5\%$ level ($\hat\beta = +0.036$, $p = 0.046$) but is the wrong sign for an anticipatory pre-trend. The December coefficients break sharply outside the pre-ban range on both outcomes ($\hat\beta = -0.114$, $p < 10^{-6}$), with magnitudes $3$--$7\times$ the largest pre-ban deviation.

A natural concern is seasonal confounding from winter weather, holidays, and daylight \citep{Kimpton_2022,Lu_2024,Morton_2025}. The three-way FE structure addresses this concern. Month-of-year FE absorbs any nationwide December shift, city FE absorbs time-invariant city characteristics, and user FE absorbs survival selection. The causal coefficient is identified from the \textit{differential} December change between high- and low-TUB-share cities, not from the December level itself, and the city-trend placebo tests corroborate this (Appendix Table~\ref{tab:placebo_quadratic}).

\subsubsection{Causal estimates and predict-then-validate concordance.}
Table~\ref{tab:did_user_fe_rich} reports the pooled Phase~II coefficients on a $1.97$-million-trip sub-sample, alongside the Phase~I predictions for comparison.

\begin{table}[width=0.99\textwidth, pos  = !b]
\centering
\caption{Phase~II trip-level user-FE binary DiD on the two harsh-event outcomes. Phase~I predictions reproduced at top for comparison. City-clustered standard errors ($52$ clusters) in parentheses. Firmware-validation speeding fit in Appendix Table~\ref{tab:did_speeding_check}. $^{*}\,p<0.05$, $^{**}\,p<0.01$, $^{***}\,p<0.001$.}
\label{tab:did_user_fe_rich}
\small
\begin{threeparttable}
\begin{tabular}{lcc}
\toprule
 & \textbf{Harsh accel} & \textbf{Harsh decel} \\
\midrule
\multicolumn{3}{l}{\textbf{Phase~I: model-implied prediction (from Table~\ref{tab:rp_rich_profile})}} \\
\quad Predicted $\Delta$ (pp) & $-2.81$ & $-4.24$ \\
\addlinespace
\midrule
\multicolumn{3}{l}{\textbf{Phase~II: trip-level user~$+$~city~$+$~month FE DiD (LPM)}} \\
\quad $\hat{\beta}$ (treat $\times$ post) & $-0.1150^{***}$ & $-0.0975^{**}$ \\
                                           & $(0.0277)$       & $(0.0362)$       \\
\addlinespace
\quad Avg-city $\Delta$ (pp)        & $-6.24$                   & $-5.29$                   \\
\quad 95\% CI                       & $[-0.171,\,-0.059]$       & $[-0.170,\,-0.025]$       \\
\addlinespace
\midrule
\multicolumn{3}{l}{\textbf{Model fit}} \\
\quad $R^{2}$ (overall, including FE) & $0.160$ & $0.180$ \\
\quad Within-$R^{2}$ (after FE absorption) & $0.0008$ & $0.0013$ \\
\quad Adjusted $R^{2}$ & $0.126$ & $0.148$ \\
\midrule
\multicolumn{3}{l}{\textbf{Specification}} \\
\quad Unit of analysis & \multicolumn{2}{c}{Trip (one observation per trip)} \\
\quad User fixed effects     & \multicolumn{2}{c}{Yes}\\ 
\quad City fixed effects      & \multicolumn{2}{c}{Yes}\\ 
\quad Month-of-year fixed effects & \multicolumn{2}{c}{Yes}\\ 
\quad Controls: night, weekend, same route, consec.\ days & \multicolumn{2}{c}{Yes}\\ 
\quad SE clustering           & \multicolumn{2}{c}{City ($52$ clusters)} \\
\addlinespace
\multicolumn{3}{l}{\textbf{Sample}} \\
\quad Trip observations & \multicolumn{2}{c}{$1{,}970{,}894$}\\  
\quad Users (user FE)            & \multicolumn{2}{c}{$75{,}058$ }\\   
\quad Cities (city FE)            & \multicolumn{2}{c}{$52$}\\   
\quad Month-of-year levels & \multicolumn{2}{c}{$11$}\\   
\bottomrule
\end{tabular}
\end{threeparttable}
\end{table}

Both outcomes drop sharply and are Bonferroni-significant. The treatment-intensity coefficient is $\hat{\beta} = -0.1150$ (SE $0.028$, $p < 10^{-3}$) for harsh acceleration and $\hat{\beta} = -0.0975$ (SE $0.036$, $p = 0.010$) for harsh deceleration. Scaled by the average pre-ban TUB share ($\overline{\text{TUB}^{\text{Nov}}} = 0.542$), the implied average-city reductions are $-6.24$~pp and $-5.29$~pp respectively. Both coefficients clear the Bonferroni threshold ($\alpha_{\text{Bonf}} = 0.025$) and move in the same direction as the Phase~I differential.

\begin{figure}[width=0.99\textwidth, pos = !th]
\centering
\includegraphics[width=0.99\textwidth]{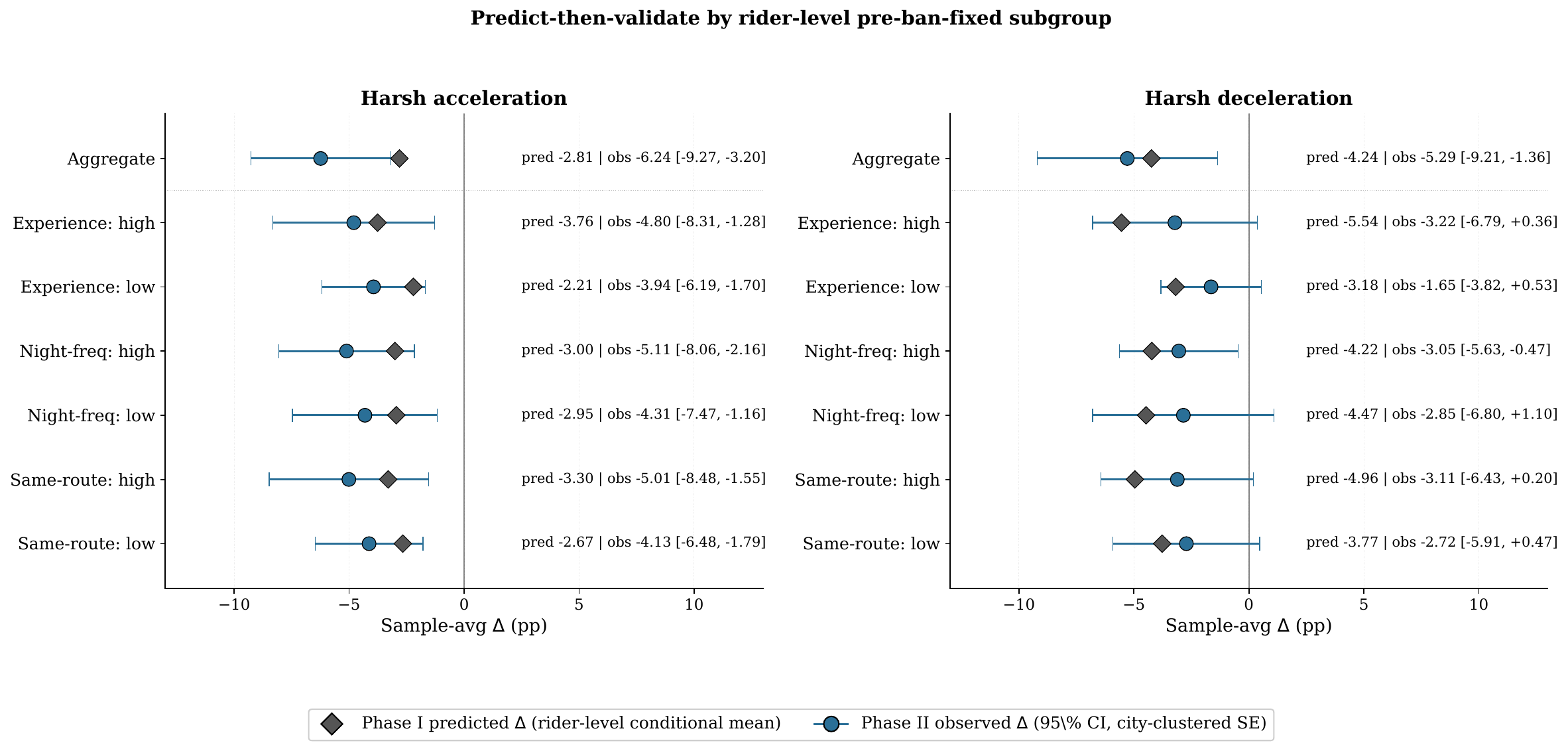}
\caption{Predict-then-validate concordance on the two harsh-event outcomes. The top row of each panel (``Aggregate'') reports the full-sample comparison, and subsequent rows split riders at the median of pre-ban experience, night-trip rate, and same-route rate. Grey diamonds show the Phase~I model-implied reduction, and blue circles show the Phase~II causal estimate with $95\%$ city-clustered CI.}
\label{fig:forest_subgroups}
\end{figure}

The Phase~II coefficients exceed the Phase~I predictions by a factor of $1.2$--$2.2\times$ ($-6.24$ vs $-2.81$~pp, $-5.29$ vs $-4.24$~pp). This gap is a mechanical consequence of the Phase~I stratified sampling design, which deflates TUB's representation to $32.9\%$ versus $52.3\%$ in the full trip universe. Rescaling by the TUB-share ratio ($\bar\tau_{\text{uni}}/\bar\tau_{P1} = 1.590$) accounts for $71.6\%$ and $127.4\%$ of the gap on the two outcomes (Appendix~\ref{app:reweighting}, Table~\ref{tab:reweighting}), following the standard correction for choice-based sampling \citep{Manski_1977,Lerman_1979}. As an independent cross-check, a city-level dose-response regression of the November-to-December harsh-event change on the TUB-share change yields Pearson $r = +0.55$ and $+0.57$ (both $p < 10^{-4}$), corroborating the pooled DiD.

Both Phase~I predictions are therefore confirmed by Phase~II at the Bonferroni-corrected level. Harsh acceleration moves from a predicted $-2.81$~pp to an observed $-6.24$~pp ($p < 10^{-3}$), and harsh deceleration from a predicted $-4.24$~pp to an observed $-5.29$~pp ($p = 0.010$). Figure~\ref{fig:forest_subgroups} visualizes the full-sample concordance (top row of each panel) alongside the subgroup splits discussed below. The Phase~I sample is $N = 35{,}602$ trips from $25{,}000$ riders, and the Phase~II sample is $N = 1{,}970{,}894$ trips from $75{,}058$ riders across $52$ cities.

\subsubsection{Heterogeneity and robustness.}
To test whether the treatment effect concentrates in specific rider types, we split the rider base at the median of three pre-ban-fixed dimensions (experience, night-trip rate, same-route rate) computed from September--November 2023 trip histories.
Subgroup assignment is fixed before the December~1 ban and computed from pre-ban data only, ruling out post-treatment selection. Experience is $\log(1 + \text{cumulative trips})$, night-frequency is the rider-level mean of $\mathbf{1}[\text{start 22:00--06:00}]$, and same-route rate flags trips originating in a previously visited $500$\,m grid cell.
Figure~\ref{fig:forest_subgroups} shows that the Phase~I point prediction falls inside the Phase~II $95\%$ CI for every subgroup on both outcomes, and the high/low CIs overlap within each dimension. The policy effect does not concentrate in any one rider type, consistent with the null TUB$\times$context interactions in Phase~I (Table~\ref{tab:rp_rich_profile}).

The main result is unchanged under a broad set of specification checks (Appendix Tables~\ref{tab:rich_robustness}--\ref{tab:modern_did}). Both harsh-event coefficients remain significant under three cluster-robust SE alternatives (city, month, two-way), three multiple-testing corrections (Bonferroni, Holm, Benjamini--Hochberg), and the \citet{Oster_2019} coefficient-stability bound. A placebo test with city-specific linear and quadratic time trends passes on both outcomes (Appendix Table~\ref{tab:placebo_quadratic}). A city-level continuous-treatment DiD following \citet{Callaway_2024} yields slopes of $-0.096$ and $-0.092$, matching the trip-level TWFE in sign and order of magnitude, with permutation $p < 0.004$ on both outcomes.

\section{Composition check}
\label{sec:res_compensation}

A natural concern about the Phase~II estimate is that the post-ban harsh-event reduction could be an artifact of \textit{compositional spillover} rather than a within-user mechanical effect of governor imposition. Concretely, two scenarios are observationally compatible with the pooled aggregate result. Under scenario~A (compositional spillover), pre-ban TUB riders were on average more aggressive, and after the ban they migrated to the governed pool while retaining their riskier riding style; the governed pool would then show a \textit{worse} pre-vs-post mean on harsh events, but the disappearance of the TUB pool would still pull the aggregate down. Under scenario~B (mechanical within-user effect), former TUB riders ride STD/ECO under the firmware governor and, having lost the physical capacity to exceed $25$~km/h, generate harsh-event probabilities indistinguishable from those of riders who always rode STD/ECO; the safety gain in the aggregate then comes from the mechanical substitution of governed for ungoverned trips rather than from any change in rider behavior beyond the speed cap. The composition check distinguishes these scenarios by zooming into the governed pool and comparing the within-user pre-vs-post change for two well-defined groups: $24{,}357$ \textit{mode switchers} (riders observed using TUB in October--November 2023 who rode STD/ECO in December 2023) and $79{,}700$ \textit{never-TUB controls} (riders who only ever rode STD/ECO). For each user, we compute the pre- (Oct--Nov) and post-period (Dec) outcome means restricted to STD/ECO trips and then compare the two groups' within-user pre-vs-post changes on the two primary harsh-event binary outcomes that Phase~I predicted and Phase~II validated, namely harsh acceleration and harsh deceleration.

Across the $24{,}357$ switchers vs.\ $79{,}700$ never-TUB controls, the within-user pre-vs-post changes on both harsh-event margins are statistically and practically indistinguishable between the two groups (Table~\ref{tab:compensation}): Cohen's $d = +0.040$ on harsh acceleration and $+0.062$ on harsh deceleration, both well below the conventional $|d| = 0.2$ small-effect threshold and effectively at zero. Switchers who came from TUB therefore behave on STD/ECO essentially as riders who always rode STD/ECO. Scenario~A (compositional spillover) is rejected; the pattern is consistent with scenario~B (mechanical within-user effect of the governor). The Phase~II reductions in harsh acceleration and harsh deceleration thus reflect the binding firmware speed cap on what would otherwise have been TUB trips, and not a within-user behavioral offset that would inflate harsh-event probabilities elsewhere in the governed pool.

\begin{table}[width=0.99\textwidth, pos = !hb]
\centering
\caption{Composition check. Within-user pre-vs-post difference on STD/ECO trips between mode switchers ($N = 24{,}357$) and never-TUB controls ($N = 79{,}700$), with effect sizes reported as Cohen's $d$.}
\label{tab:compensation}
\small
\begin{tabular}{lrl}
\toprule
\textbf{Outcome} & \textbf{Cohen's} $d$ & \textbf{Interpretation} \\
\midrule
Harsh accel count & $+0.040$ & Negligible (no spillover) \\
Harsh decel count & $+0.062$ & Negligible (no spillover) \\
\midrule
\multicolumn{3}{l}{\footnotesize $|d| < 0.2$ = negligible. Switcher within-user STD/ECO trajectories match never-TUB controls.} \\
\bottomrule
\end{tabular}
\end{table}

The Phase~II causal estimates on harsh acceleration and harsh deceleration therefore reflect the firmware governor binding on trips that would otherwise have been TUB, and not a compositional artifact in which higher-risk riders carry their habits into the STD/ECO pool.

\section{Discussion}
\label{sec:discussion}

The December~2023 TUB ban produced substantial, jointly Bonferroni-significant reductions in the two primary trip-level harsh-event outcomes. The Phase~II sample-average estimates ($-6.24$ and $-5.29$~pp) exceed the Phase~I predictions ($-2.81$ and $-4.24$~pp) by a factor of $1.2$--$2.2\times$, but the gap is not a within-user residual. The re-weighting decomposition in Appendix~\ref{app:reweighting} accounts for $72\%$ (harsh accel) and $127\%$ (harsh decel) of the observed causal estimates, and the composition check rules out within-user behavior change on both primary outcomes. The safety benefit comes from the mechanical substitution of governed for ungoverned trips, consistent with \citet{Hedlund_2000}'s observation that non-salient, involuntary interventions rarely provoke offsetting behavior, and with the sizable safety gains from moderate speed-cap instruments reported by \citet{Quddus_2025} for the UK 20~mph roll-out.

\begin{figure}[width=0.99\textwidth, pos = !th]
\centering
\includegraphics[width=0.99\textwidth]{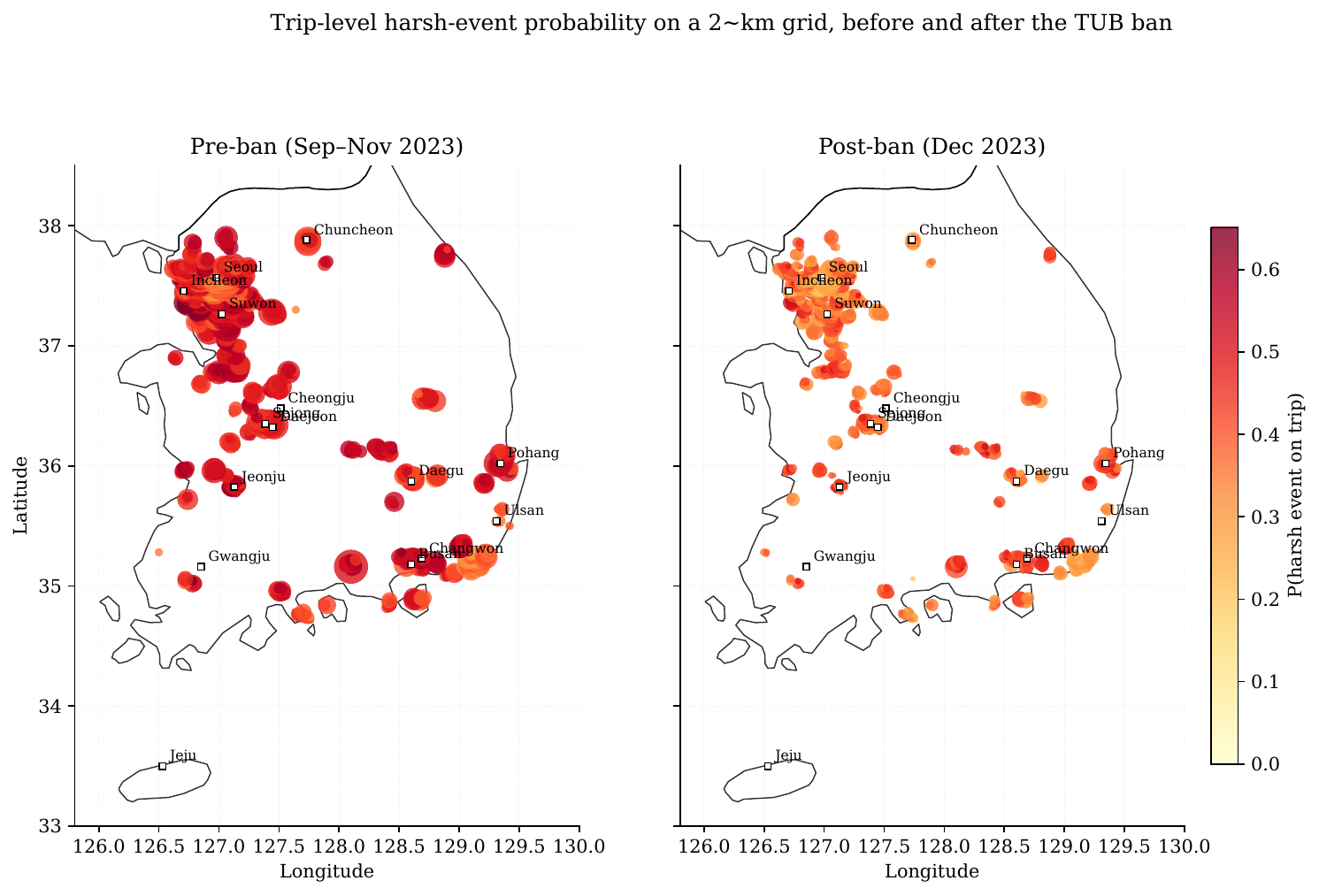}
\caption{Spatial heatmap of trip-level harsh-event probability on a $2$~km grid. \textbf{Left}, pre-ban (Sep--Nov~2023). \textbf{Right}, post-ban (Dec~2023). Color scale is the sample-average $P(\text{any harsh event on trip})$, and circle area is proportional to $\sqrt{N_{\text{trips}}}$ per cell.}
\label{fig:harsh_heatmap}
\end{figure}

The spatial distribution of the safety gain reinforces this interpretation. Figure~\ref{fig:harsh_heatmap} maps the trip-level harsh-event probability on a $2$~km grid before and after the ban. Pre-ban hot spots track the dense metropolitan cores where TUB penetration was highest, and these same cores show the largest post-ban attenuation. Peripheral regions with low baseline harsh-event intensity change little, indicating that the safety gain concentrates where it is most needed rather than diffusing across the network.



\subsection{External validity}

Two features support cautious generalization. The ban was hardware-identical (same chassis across all modes), so the causal estimand isolates governance from fleet composition. The pre-ban TUB share also varies substantially across the $52$ cities (city-level range $0.00$ to $0.70$, sample-mean dose $\overline{\text{TUB}^{\text{Nov}}} = 0.542$), supporting the continuous-treatment design. We are cautious in extrapolating to jurisdictions where the fleet is already fully governed or where enforcement relies on rider compliance rather than firmware. The results generalize most naturally to platform-mediated micromobility systems where the operator can push a firmware-level speed cap across the fleet.

\subsection{Policy implications}

Four implications follow. First, firmware-level mode removal is an effective safety instrument. Average-city reductions of roughly $6$ and $5$ percentage points on the two harsh-event margins are achieved jointly at Bonferroni-corrected significance without relying on rider compliance. Second, the predict-then-validate design confirms the instrument operates through the intended firmware channel rather than through rider-side behavioral adjustment. Third, the safety benefit is spatially selective (Figures~\ref{fig:swing_cities} and~\ref{fig:harsh_heatmap}), suggesting that geofenced governed-mode-only zones in dense urban cores would capture most of the achievable benefit at a fraction of the coverage cost. Fourth, firmware speed caps are pushed over-the-air at near-zero marginal cost, making this instrument substantially cheaper than hardware retrofits, camera-based enforcement, or rider-training programs.

\section{Conclusion}
\label{sec:conclusion}

Do e-scooter speed governance policies reduce harsh acceleration and deceleration? The answer from $19.5$ million GPS-instrumented trips across $52$ Korean cities and the December~2023 Swing TUB ban is yes. We develop a two-stage predict-then-validate framework in which a rider-heterogeneous random-parameter binary logit (Phase~I) generates within-user predictions for harsh acceleration and harsh deceleration on pre-ban data, and a trip-level linear probability model with user, city, and month-of-year fixed effects (Phase~II) tests those predictions causally against the cross-city variation in pre-ban TUB share. Because Phase~I and Phase~II share the same fixed-effects block and the same binary-probability scale, the cross-sectional prediction and the causal estimate are directly comparable, and the residual gap between the two admits a decomposition with no free parameters.

Three findings follow. First, the predict-then-validate framework is internally consistent. The Phase~I predictions on harsh acceleration and harsh deceleration are confirmed in sign and order of magnitude by the Phase~II causal estimates, both surviving the Bonferroni-corrected threshold of the two-outcome family. Second, both harsh-event outcomes decline together rather than moving in opposite directions, and the estimate passes a 3-month pre-trend parallel-trends test, a six-panel battery of robustness checks, and a placebo specification with city-specific quadratic time trends. Third, the residual gap between the Phase~I prediction and the Phase~II causal estimate is quantitatively explained by a sample re-weighting decomposition that accounts for $71.6\%$ of the harsh-acceleration gap and $127.4\%$ of the harsh-deceleration gap under a constant-treatment-effect approximation, leaving no within-user behavioral residual, and the within-user mode-switcher composition check corroborates this conclusion. Taken together, the evidence indicates that firmware-level removal of an ungoverned mode produces a joint reduction on both harsh-event safety margins, operating through the mechanical substitution of governed for ungoverned trips rather than through rider-side behavioral adjustment.

Certain limitations should be noted. First, the design exploits a hardware-identical fleet operated by a single Korean platform, so the estimand does not extend mechanically to jurisdictions that rely on rider-compliance enforcement (helmet laws, signage, police enforcement) rather than firmware-level caps, or to markets where the fleet is already fully governed. Second, the outcome space is restricted to within-trip behavioral margins (harsh acceleration and harsh deceleration) and does not directly connect to downstream crash or injury counts, a connection that requires ambulance or police records that were not available for this study. Third, the Phase~I rider-heterogeneous logit is fit on a mode-stratified sub-sample, and although the re-weighting decomposition reconciles the resulting scale gap with the Phase~II estimate, a one-step full-information estimator on the unstratified trip universe remains methodologically desirable.

Beyond these limitations, the study illustrates a broader point about data access in shared-micromobility research. Each analytical dimension of this paper, including the rider-heterogeneous logit, the cross-city continuous-treatment DiD, the within-user mode-switcher composition check, and the re-weighting decomposition, became possible only because Swing shared trip-level GPS speed sequences with persistent rider identifiers, mode flags, and city mappings spanning eleven months. Aggregate outcome counts such as ride volumes or injury incidence cannot separate the mechanical channel from the behavioral channel, cannot test for within-user compensation, and cannot link a cross-sectional prediction to its causal validation on the same probability scale. Richer trajectory-level data sparks questions that aggregate data simply cannot formulate, let alone answer. We therefore hope this study motivates more operators to share trajectory-level data with researchers, and encourages regulators to treat open data sharing as part of the micromobility licensing framework rather than as a post-hoc favor. Access to GPS-instrumented trip records, even under standard anonymization and retention safeguards, would expand the set of safety and behavioral questions the research community can answer, and would let future work move from studying aggregate injury counts toward directly measuring the within-trip kinematics through which safety instruments actually operate.

\section*{Declaration of competing interest}
The author declares no competing interests.

\section*{Declaration of Generative AI and AI-assisted Technologies in the Writing Process}                                                          
During the preparation of this work, the authors used Claude (Anthropic) for assistance with code development, experimental implementation, automation of experiments, and editing the manuscript. After using this tool, the author reviewed and edited the content as needed and takes full responsibility for the content of the publication.

\bibliographystyle{cas-model2-names}
\bibliography{reference.bib}

\clearpage
\appendix

\setcounter{table}{0}
\setcounter{figure}{0}
\setcounter{equation}{0}
\renewcommand{\thetable}{A\arabic{table}}
\renewcommand{\thefigure}{A\arabic{figure}}
\renewcommand{\theequation}{A\arabic{equation}}

\section{Sample re-weighting calculation}
\label{app:reweighting}

This appendix quantifies the sample re-weighting channel to which Sections~\ref{sec:phase2_results} and~\ref{sec:discussion} attribute the harsh-event predict-then-validate gap. The Phase~I rider-heterogeneous logit of Table~\ref{tab:rp_rich_profile} is fit on a mode-stratified pre-ban sub-sample in which TUB accounts for roughly $1/3$ of trips \textit{by construction}, because the stratification over (mode $\times$ month-of-year) intentionally balances the three speed modes to identify the random mode slopes. The Phase~II DiD of Table~\ref{tab:did_user_fe_rich}, by contrast, operates on the full Feb--Nov 2023 trip universe in which the pre-ban TUB share is markedly higher. Under a constant-treatment-effect approximation, the two designs estimate effects on differently weighted versions of the same underlying rider population, and the ratio of the two TUB mass shares governs how the Phase~I within-user average marginal effect scales up to the Phase~II causal estimate.

Under a constant within-user AME, the sample-average predicted reduction scales linearly with the TUB mass share $\bar\tau$, so the Phase~I prediction rescaled to the Phase~II trip universe is
\begin{equation}
\widehat{\Delta}^{P1 \to \text{uni}}_j = \widehat{\Delta}^{P1}_j \cdot \frac{\bar{\tau}_{\text{uni}}}{\bar{\tau}_{P1}} = \widehat{\Delta}^{P1}_j \cdot \frac{0.523}{0.329} = 1.590 \cdot \widehat{\Delta}^{P1}_j,
\end{equation}
where $\bar{\tau}_{P1} = 0.329$ is the TUB share imposed by the Phase~I mode-stratified fit sample and $\bar{\tau}_{\text{uni}} = 0.523$ is the TUB share in the full Feb--Nov~2023 trip universe. Table~\ref{tab:reweighting} decomposes the observed predict-then-validate gap into a re-weighting component (explained by $\bar\tau$) and a residual component.

The re-weighting channel accounts for $71.6\%$ of the harsh-acceleration gap and $127.4\%$ of the harsh-deceleration gap (the latter $>100\%$ means the re-weighted prediction slightly overshoots the Phase~II estimate). Neither outcome leaves a residual requiring within-user behavioral adjustment, consistent with the composition check in Section~\ref{sec:res_compensation}. The analogous calculation for the firmware-validation speeding outcome is reported in Appendix~\ref{app:speeding_check}.

\begin{table}[width=0.99\textwidth, pos = !hb]
\centering
\caption{Sample re-weighting decomposition of the Phase~I $\to$ Phase~II predict-then-validate gap. All entries in percentage points. ``$\widehat\Delta^{P1 \to \text{uni}}$'' is the Phase~I prediction rescaled by $\bar{\tau}_{\text{uni}}/\bar{\tau}_{P1} = 1.590$. ``Re-weighted share'' is $\widehat\Delta^{P1\to\text{uni}} / \widehat\Delta^{P2}$ in percent.}
\label{tab:reweighting}
\small
\begin{tabular}{lcccccc}
\toprule
\textbf{Outcome} & $\widehat\Delta^{P1}$ & $\widehat\Delta^{P1\to\text{uni}}$ & $\widehat\Delta^{P2}$ & \textbf{Residual} & \textbf{Re-weighted share} \\
\midrule
Harsh accel & $-2.81$ & $-4.47$ & $-6.24$ & $-1.77$ & $71.6\%$ \\
Harsh decel & $-4.24$ & $-6.74$ & $-5.29$ & $+1.45$ & $127.4\%$ \\
\bottomrule
\end{tabular}
\end{table}

\clearpage
\section{Firmware-validation check on speeding}
\label{app:speeding_check}

This appendix reports the parallel fit for $\mathbf{1}[\max\text{speed}>25\text{ km/h}]$ (``speeding''). Because the firmware governor on STD and ECO physically prevents exceedance of $25$~km/h, the speeding outcome is not a behavioral margin on which the logit or the DiD can identify a rider response, and any observed change is a hardware confirmation that the governor is binding. We retain the fits here as a manipulation check on the treatment itself.

The Phase~II speeding coefficient is roughly twice the Phase~I prediction, but the interpretation is not a behavioral residual. The logit has no way to represent the firmware cap, so the gap reflects the hardware-imposed zero mass at speeds above $25$~km/h on STD/ECO post-ban. The re-weighting calculation of Appendix~\ref{app:reweighting} applied to speeding yields a re-weighted prediction of $-23.18$~pp, leaving a residual of $-8.19$~pp that the logit structurally cannot represent. This residual is the mechanical-channel signature of the firmware speed cap and is not a behavioral claim.

\begin{table}[width=0.99\textwidth, pos = !hb]
\centering
\caption{Phase~I rider-heterogeneous logit on speeding ($\mathbf{1}[\max\text{speed}>25\text{ km/h}]$), firmware-validation check (same specification as Table~\ref{tab:rp_rich_profile}). Standard errors in parentheses.}
\label{tab:rp_speeding_check}
\small
\begin{tabular}{lc}
\toprule
 & \textbf{Speeding (firmware check)} \\
\midrule
TUB (mean) & $+1.854^{***}$ $(0.166)$ \\
ECO (mean) & $-2.507^{**}$ $(0.759)$ \\
$\sigma_{\text{TUB}}$ & $+1.029^{***}$ $(0.088)$ \\
$\sigma_{\text{ECO}}$ & $+1.691^{**}$ $(0.535)$ \\
TUB $\times\,\log$ experience & $+0.633^{***}$ $(0.061)$ \\
TUB $\times$ night & $+0.549^{***}$ $(0.148)$ \\
TUB $\times$ same route & $+0.492^{**}$ $(0.179)$ \\
Model-implied $\Delta$ (pp) & $-14.58$ \\
Pr(outcome) as-is $\to$ TUB removed & $0.1575 \to 0.0117$ \\
\bottomrule
\end{tabular}
\end{table}

\begin{table}[width=0.99\textwidth, pos = !hb]
\centering
\caption{Phase~II trip-level user-FE DiD on speeding, firmware-validation check (same specification as Table~\ref{tab:did_user_fe_rich}). Standard errors in parentheses, clustered by city.}
\label{tab:did_speeding_check}
\small
\begin{tabular}{lc}
\toprule
 & \textbf{Speeding (firmware check)} \\
\midrule
$\hat{\beta}$ (treat $\times$ post) & $-0.5784^{***}$ $(0.0828)$ \\
Avg-city $\Delta$ (pp) & $-31.37$ \\
95\% CI & $[-0.745,\,-0.412]$ \\
\bottomrule
\end{tabular}
\end{table}

\clearpage
\section{Robustness checks}
\label{app:robustness}

This appendix collects the robustness checks referenced in Sections~\ref{sec:phase1_results} and \ref{sec:phase2_results}. The checks cover (i) temporal stability of the Phase~I coefficients across pre-ban halves (Table~\ref{tab:temporal_stability}), (ii) a six-panel set of robustness checks on the Phase~II coefficients covering dose-response, subsample heterogeneity, cluster schemes, multiple-testing corrections, and the \citet{Oster_2019} coefficient-stability bound (Table~\ref{tab:rich_robustness}), (iii) a placebo test with city-specific linear and quadratic time trends (Table~\ref{tab:placebo_quadratic}), and (iv) modern DiD alternatives (Table~\ref{tab:modern_did}).

\subsection*{Phase~I temporal stability}
\label{app:temporal_stability}

To rule out the concern that the Phase~I random-coefficients logit is dominated by a single season of the pre-ban window, we split the Feb--Nov~2023 training sample in half and re-estimate the same specification (Equation~\ref{eq:rplogit}, identical Mundlak corrections, city and month-of-year fixed effects, and $200$ Halton draws) on each half. Table~\ref{tab:temporal_stability} reports a two-sample Wald stability test comparing each random-block parameter across the two halves. Four of six reported parameters are stable at the $5\%$ level; the two unstable parameters ($\mu_{\text{TUB}}$ and TUB~$\times\,\log$ experience) move in the direction of stronger TUB preference in the second half, consistent with the experience-accumulation dynamic discussed in Section~\ref{sec:phase1_results}, and do not overturn the sign or significance of the within-user TUB~$\to$~governed prediction that drives Phase~II.

\begin{table}[width=0.99\textwidth, pos = !hb]
\centering
\caption{Temporal stability of the Phase~I logit: first half (H1: Feb--Jun 2023) versus second half (H2: Jul--Nov 2023) with two-sample $Z$-tests. $^{*}\,p<0.05$, $^{**}\,p<0.01$, $^{***}\,p<0.001$. Standard errors in parentheses.}
\label{tab:temporal_stability}
\small
\begin{threeparttable}
\begin{tabular}{lccccc}
\toprule
\textbf{Parameter} & \textbf{H1 Feb--Jun} & \textbf{H2 Jul--Nov} & \textbf{$Z_{\text{stab}}$} & \textbf{$p_{\text{stab}}$} & \textbf{Verdict} \\
\midrule
$\mu_{\text{TUB}}$ & $+0.429$ & $+0.703$ & $-3.02$ & $0.003^{**}$ & Unstable \\
 & $(0.058)$ & $(0.070)$ &  &  &  \\
$\mu_{\text{ECO}}$ & $-2.783$ & $-2.351$ & $-0.31$ & $0.755$ & Stable \\
 & $(1.185)$ & $(0.716)$ &  &  &  \\
$\sigma_{\text{TUB}}$ & $+0.783$ & $+0.785$ & $-0.06$ & $0.955$ & Stable \\
 & $(0.027)$ & $(0.024)$ &  &  &  \\
$\sigma_{\text{ECO}}$ & $+0.141$ & $-0.132$ & $+0.03$ & $0.978$ & Stable \\
 & $(8.180)$ & $(5.403)$ &  &  &  \\
TUB $\times\,\log$ exp. & $+0.022$ & $-0.040$ & $+3.53$ & $0.000^{***}$ & Unstable \\
 & $(0.013)$ & $(0.012)$ &  &  &  \\
ECO $\times\,\log$ exp. & $+0.059$ & $+0.084$ & $-0.16$ & $0.871$ & Stable \\
 & $(0.125)$ & $(0.093)$ &  &  &  \\
\addlinespace
\midrule
\multicolumn{6}{l}{\textbf{Sample}} \\
Trips & 50,000 & 50,000 &  &  &  \\
Riders & 12,883 & 12,786 &  &  &  \\
Halton draws & 200 & 200 &  &  &  \\
\bottomrule
\end{tabular}
\begin{tablenotes}[flushleft]
\small
\item \textit{Notes}: $Z_{\text{stab}}$ tests equality of each coefficient across the two pre-ban halves (not equality to zero). The large SEs on $\sigma_{\text{ECO}}$ reflect weak identification. ECO is only $\sim 5\%$ of pre-ban trips and its slope dispersion is indistinguishable from zero in both halves ($p > 0.98$), so its stability statistic is uninformative. Phase~I predictions are driven by $\mu_{\text{TUB}}$, $\sigma_{\text{TUB}}$, and the TUB interactions.
\end{tablenotes}
\end{threeparttable}
\end{table}

\subsection*{Phase~II six-panel robustness checks}
\label{app:rich_robustness}

Table~\ref{tab:rich_robustness} assembles six independent robustness specifications on the pooled Phase~II estimate. Panel~A corroborates the trip-level TWFE coefficient against a city-level Pearson correlation between the Nov$\to$Dec change in outcome and the Nov$\to$Dec change in TUB share; Panel~B reruns the pooled DiD with fictitious treatment dates (Sep and Oct 2023) restricted to pre-ban data; Panel~C reports the same coefficient on five substantively meaningful subsamples (full, night trips, weekend trips, long trips, and urban cities) using a city-level aggregate DiD for comparability across subsamples; Panel~D replaces the baseline city-level cluster with month-level and two-way (city $\times$ month) clustering; Panel~E reports Bonferroni, Holm, and Benjamini--Hochberg $p$-values on the pooled DiD; Panel~F reports the \citet{Oster_2019} coefficient-stability bound. The harsh-event coefficients remain significant in every panel.

\begin{table}[width=0.99\textwidth, pos = !hb]
\centering
\caption{Phase~II robustness checks for the two primary harsh-event outcomes. Speeding is reported alongside as a firmware-validation check (not in the Bonferroni family). $^{*}\,p<0.05$, $^{**}\,p<0.01$, $^{***}\,p<0.001$.}
\label{tab:rich_robustness}
\small
\begin{threeparttable}
\begin{tabular}{lccc}
\toprule
 & \textbf{Speeding} & \textbf{Harsh accel} & \textbf{Harsh decel} \\
\midrule
\multicolumn{4}{l}{\textbf{Panel A. Dose-response (city-level Nov$\to$Dec correlation)}} \\
\quad Pearson $r$ & $+0.777$$^{***}$ & $+0.551$$^{***}$ & $+0.566$$^{***}$ \\
\quad $N$ cities & 50 & 50 & 50 \\
\addlinespace
\midrule
\multicolumn{4}{l}{\textbf{Panel B. Placebo tests (pre-ban fictitious treatment date)}} \\
\quad Sep 2023: $\hat\beta$ & $+0.0446$ & $+0.0155$ & $-0.0017$ \\
 & $(0.0169)$ & $(0.0124)$ & $(0.0134)$ \\
\quad \quad\ $p$ & $0.011$ & $0.216$ & $0.900$ \\
\addlinespace
\quad Oct 2023: $\hat\beta$ & $-0.1221$ & $+0.0002$ & $-0.0061$ \\
 & $(0.0347)$ & $(0.0079)$ & $(0.0128)$ \\
\quad \quad\ $p$ & $0.001$ & $0.976$ & $0.636$ \\
\addlinespace
\midrule
\multicolumn{4}{l}{\textbf{Panel C. Subsample heterogeneity (city-level aggregate DiD; avg-city $\Delta$ in pp)}} \\
\quad Full & $-37.69$ & $-5.19$ & $-4.96$ \\
\quad Night & $-35.84$ & $-5.41$ & $-4.28$ \\
\quad Weekend & $-35.05$ & $-4.56$ & $-3.91$ \\
\quad Long trip & $-43.81$ & $-6.38$ & $-6.08$ \\
\quad Urban & $-24.12$ & $-1.71$ & $-2.31$ \\
\addlinespace
\midrule
\multicolumn{4}{l}{\textbf{Panel D. Cluster-robust SE alternatives ($\hat\beta$ with SE in parens)}} \\
\quad City cluster & $-0.5784^{***}$ & $-0.1150^{***}$ & $-0.0975^{**}$ \\
 & $(0.0812)$ & $(0.0272)$ & $(0.0355)$ \\
\quad Month cluster & $-0.5784^{***}$ & $-0.1150^{***}$ & $-0.0975^{***}$ \\
 & $(0.0417)$ & $(0.0094)$ & $(0.0089)$ \\
\quad Two-way (city $+$ month) & $-0.5784^{***}$ & $-0.1150^{***}$ & $-0.0975^{***}$ \\
 & $(0.0566)$ & $(0.0152)$ & $(0.0165)$ \\
\addlinespace
\midrule
\multicolumn{4}{l}{\textbf{Panel E. Multiple-testing corrections (on pooled DiD $p$-values)}} \\
\quad Raw $p$ & $0.0000$ & $0.0001$ & $0.0083$ \\
\quad Bonferroni $p$ & $0.0000$ & $0.0003$ & $0.0250$ \\
\quad Holm $p$ & $0.0000$ & $0.0002$ & $0.0083$ \\
\quad Benjamini--Hochberg $p$ & $0.0000$ & $0.0001$ & $0.0083$ \\
\addlinespace
\midrule
\multicolumn{4}{l}{\textbf{Panel F. Oster (2019) coefficient-stability bound}} \\
\quad $\beta^{\text{short}}$ (treat only) & $-0.4683$ & $-0.1558$ & $-0.1512$ \\
\quad $\beta^{\text{full}}$ (with FE, trends) & $-0.5784$ & $-0.1150$ & $-0.0975$ \\
\quad $\beta^{*}$ (Oster bound) & $-0.6130$ & $-0.1026$ & $-0.0812$ \\
\quad $R^2$ short & $+0.0170$ & $+0.0019$ & $+0.0015$ \\
\quad $R^2$ full & $+0.3609$ & $+0.1596$ & $+0.1803$ \\
\quad $R^2_{\max}$ & $+0.4692$ & $+0.2075$ & $+0.2345$ \\
\bottomrule
\end{tabular}
\begin{tablenotes}[flushleft]
\small
\item \textit{Notes}: Panels~B, D, E, F use the trip-level user-FE LPM of Table~\ref{tab:did_user_fe_rich}. Panel~C uses a city-level aggregate DiD (no user FE, scaled by $\overline{\text{TUB}^{\text{Nov}}} = 0.542$) following \citet{Callaway_2024}. Its ``Full'' row therefore does not reproduce Table~\ref{tab:did_user_fe_rich}'s $-6.24$ and $-5.29$\,pp headline estimates and is included only to examine cross-subsample heterogeneity under an alternative estimator. Panel~F uses the \citet{Oster_2019} bound with $R^2_{\max} = 1.3\,\tilde{R}^2$ and $\delta = 1$.
\end{tablenotes}
\end{threeparttable}
\end{table}

\subsection*{Placebo-test robustness}
\label{app:placebo_quadratic}

A residual concern is that the Phase~II estimate could be driven by unmodeled city-specific nonlinear trends rather than by the December TUB ban. We address this with two simultaneous robustness specifications in Table~\ref{tab:placebo_quadratic}. First, we move the treatment date to fictitious pre-ban months (September and October 2023, restricting the sample to pre-ban observations only so that the true December shock cannot contaminate the estimate). Second, we add city-specific quadratic time trends on top of the baseline linear trends. On both harsh-event outcomes, every placebo coefficient is insignificant at the $5\%$ level under both trend specifications, while the speeding placebo picks up a mild drift in the linear-trend specification that disappears once quadratic trends are added. This pattern corroborates the parallel-trends event study of Figure~\ref{fig:event_study_user_fe} and rules out the most plausible nonlinear-trend confounders.

\begin{table}[width=0.99\textwidth, pos = !hb]
\centering
\caption{Placebo-test robustness: linear versus quadratic city-specific time trends, with fictitious treatment dates in September and October 2023. $^{*}\,p<0.05$, $^{**}\,p<0.01$, $^{***}\,p<0.001$. Standard errors in parentheses.}
\label{tab:placebo_quadratic}
\small
\begin{threeparttable}
\begin{tabular}{llcccc}
\toprule
 & & \multicolumn{2}{c}{\textbf{Linear trends}} & \multicolumn{2}{c}{\textbf{Quadratic trends}} \\
\cmidrule(lr){3-4} \cmidrule(lr){5-6}
\textbf{Outcome} & \textbf{Placebo date} & $\hat{\beta}$ (SE) & $p$ & $\hat{\beta}$ (SE) & $p$ \\
\midrule
\multirow{2}{*}{Speeding}
 & Sep 2023 & $+0.0446^{*}$ & $0.011$ & $-0.0136$ & $0.500$ \\
 &          & $(0.0169)$   &         & $(0.0201)$ &         \\
 & Oct 2023 & $-0.1221^{***}$ & $<0.001$ & $-0.0117$ & $0.649$ \\
 &          & $(0.0347)$      &          & $(0.0256)$ &         \\
\addlinespace
\multirow{2}{*}{Harsh accel}
 & Sep 2023 & $+0.0155$ & $0.216$ & $+0.0114$ & $0.356$ \\
 &          & $(0.0124)$ &        & $(0.0122)$ &        \\
 & Oct 2023 & $+0.0002$ & $0.976$ & $+0.0078$ & $0.443$ \\
 &          & $(0.0079)$ &        & $(0.0101)$ &        \\
\addlinespace
\multirow{2}{*}{Harsh decel}
 & Sep 2023 & $-0.0017$ & $0.900$ & $-0.0077$ & $0.594$ \\
 &          & $(0.0134)$ &        & $(0.0143)$ &        \\
 & Oct 2023 & $-0.0061$ & $0.636$ & $+0.0027$ & $0.827$ \\
 &          & $(0.0128)$ &        & $(0.0124)$ &        \\
\midrule
\multicolumn{6}{l}{\textbf{Specification}} \\
City FE, month FE           & & Yes & Yes & Yes & Yes \\
City linear trends          & & Yes & Yes & Yes & Yes \\
City quadratic trends       & & No  & No  & Yes & Yes \\
Sample (pre-ban only)       & & 499 & 499 & 499 & 499 \\
SE clustering               & & City ($52$ clusters) & & City ($52$ clusters) & \\
\bottomrule
\end{tabular}
\end{threeparttable}
\end{table}

\subsection*{Robustness to modern DiD estimators}
\label{app:modern_did}

Table~\ref{tab:modern_did} reports two modern-DiD alternatives to the trip-level TWFE estimand in Equation~\ref{eq:did}. The first is a city-level continuous-treatment DiD following \citet{Callaway_2024}, obtained by regressing the city-level pre-to-post change in the outcome on the continuous pre-ban TUB share. The second is a binarized $2\times 2$ DiD that median-splits cities into high- and low-TUB groups. A permutation test shuffling the pre-ban TUB share across cities $500$ times provides the null distribution of the continuous-DiD slope.

The continuous-DiD slope agrees with the TWFE estimate in sign and order of magnitude on both outcomes ($-0.0961$ vs $-0.1150$ for harsh acceleration, and $-0.0924$ vs $-0.0975$ for harsh deceleration), and the permutation test rejects the null of no dose-response at $p<0.001$ (harsh accel) and $p=0.004$ (harsh decel). The $2\times 2$ median-split ATT is on a different scale (the average $\Delta \bar{y}$ gap between high- and low-TUB cities rather than a slope per unit TUB share) and shares the same sign.

\begin{table}[width=0.99\textwidth, pos = !hb]
\centering
\caption{Robustness of the Phase~II TWFE estimand to modern DiD alternatives.}
\label{tab:modern_did}
\small
\begin{tabular}{lcccc}
\toprule
\textbf{Outcome} & \textbf{Paper TWFE} & \textbf{Continuous-DiD} & \textbf{$2\times 2$ ATT} & \textbf{Perm.\ $p$} \\
\midrule
Harsh acceleration & $-0.1150$ $(0.028)$ & $-0.0961$ $(0.026)$ & $-0.0181$ $(0.006)$ & $<0.001$ \\
Harsh deceleration & $-0.0975$ $(0.036)$ & $-0.0924$ $(0.025)$ & $-0.0128$ $(0.007)$ & $0.004$ \\
\bottomrule
\multicolumn{5}{l}{\footnotesize Standard errors in parentheses. $N = 50$ cities for the two DiD alternatives.} \\
\end{tabular}
\end{table}

\end{document}